\documentclass[a4paper,11pt]{article}
\usepackage{jheppub}
\usepackage{slashed}
\usepackage{booktabs}
\usepackage{graphicx}
\usepackage{multirow}
\usepackage{physics}
\usepackage[utf8]{inputenc}
\usepackage[%
separate-uncertainty=true,%
]{siunitx}
\usepackage[table]{xcolor}
\usepackage{bbm}
\usepackage{tikz}
\usepackage[margin=false,inline=true,index=true,draft]{fixme}
\fxusetheme{colorsig}
\usepackage{graphbox}
\usepackage{frcursive}
\usepackage{mathrsfs}

\begin{document}
\title{
Tree-level unitarity constraints on heavy neutral leptons
}

\author{Kevin A.\ Urqu\'ia-Calder\'on,}
\author{Inar Timiryasov,}
\author{and Oleg Ruchayskiy}

\affiliation{Niels Bohr Institute, University of Copenhagen, Jagtvej~155A, DK-2200, Copenhagen, Denmark}
\emailAdd{kevin.urquia@nbi.ku.dk}
\emailAdd{inar.timiryasov@nbi.ku.dk}
\emailAdd{oleg.ruchayskiy@nbi.ku.dk}

\arxivnumber{2409.13412}

\abstract{
  Heavy neutral leptons (HNLs) can explain the origin of neutrino masses and oscillations over a wide range of masses. Direct experimental probes of HNLs become unfeasible for masses significantly above the electroweak scale.
  Consequently, the strongest limits arise from the non-observation of charged lepton flavor-violating processes induced by HNLs at loop level.
  Counter-intuitively, these bounds tighten  as the HNL mass increases,
  an effect that persists within the perturbative regime.

  This work explores the precise form of these bounds for HNLs with masses well beyond the electroweak scale by analyzing the full matrix of partial waves (tree-level unitarity).
  At high energies, the HNL model simplifies to a Yukawa theory, allowing unitarity constraints to be expressed in terms of the total Yukawa coupling $\abs{Y_{\mathrm{tot}}}^2$ involving HNLs, lepton doublets, and the Higgs boson.
  Processes with $J=0$ and $J=1/2$ 
  yield the well-known 
  result
  $\abs{Y_{\mathrm{tot}}}^2 \leq 8\pi$.
However, the most stringent result arises from processes with $J = 1$,
which is given by $\abs{Y_{\mathrm{tot}}}^2 \leq 4\pi(\sqrt 5 -1) = 8\pi/\varphi \approx \num{15.533}$, where $\varphi$ is the Golden ratio.
These results remain valid provided that the Yukawa matrix has rank 1, a condition approximately satisfied in models with two or three HNLs, with large mixing angles, and radiactively small neutrino masses.
Finally, we determine the maximum mass that an HNL can have in the type-I seesaw model while remaining the sole source of neutrino masses.}

\maketitle
\section{Introduction}
The discovery of neutrino oscillations remains to this day the only laboratory signal that deviates from Standard Model (SM) predictions \cite{Super-Kamiokande:1998kpq, KamLAND:2003gfh, SNO:2003bmh}. 
The simplest way to account for neutrino oscillations is by including neutrino mass terms. 
Many Standard Model extensions  accommodate neutrino masses, such as the type I seesaw \cite{Minkowski:1977sc, Yanagida:1979as, Glashow:1979nm, Schechter:1981cv} and its multiple incarnations \cite{Mohapatra:1986bd, Akhmedov:1995ip}, the type II seesaw \cite{Cheng:1980qt}, type III seesaw \cite{Foot:1988aq}, loop generated models \cite{Zee:1980ai, Babu:1988ig, Babu:1988ki, Ma:2006km}, and non-minimal gauge extensions of the SM \cite{Gell-Mann:1979vob,Mohapatra:1979ia}.

In this paper, we focus on the type-I seesaw model.
This model introduces a set of electrically neutral leptons, $N_i$, referred to as Heavy Neutral Leptons (HNLs) also known as sterile neutrinos or right-handed neutrinos to the SM \cite{Abdullahi:2022jlv}.
HNLs interact only through weak-like interactions, additionally suppressed by a small mixing angle $\abs{\Theta_{\alpha i}} \ll 1$.\footnote{See Section~\ref{sec:theoretical_preliminaries} that specifies our notations.} 
The minimal number of HNLs that are needed to properly account for neutrino oscillations is two. The addition of these two HNLs can also provide a mechanism for the generation of the asymmetry between matter and antimatter in the universe, and a third, much lighter, HNL can also be a dark matter candidate \cite{Asaka:2005an, Asaka:2005pn, Boyarsky:2009ix,Boyarsky:2018tvu, Klaric:2020phc}. 

\begin{figure}[!t]
    \centering
    \includegraphics[align = c, width=0.5\linewidth]{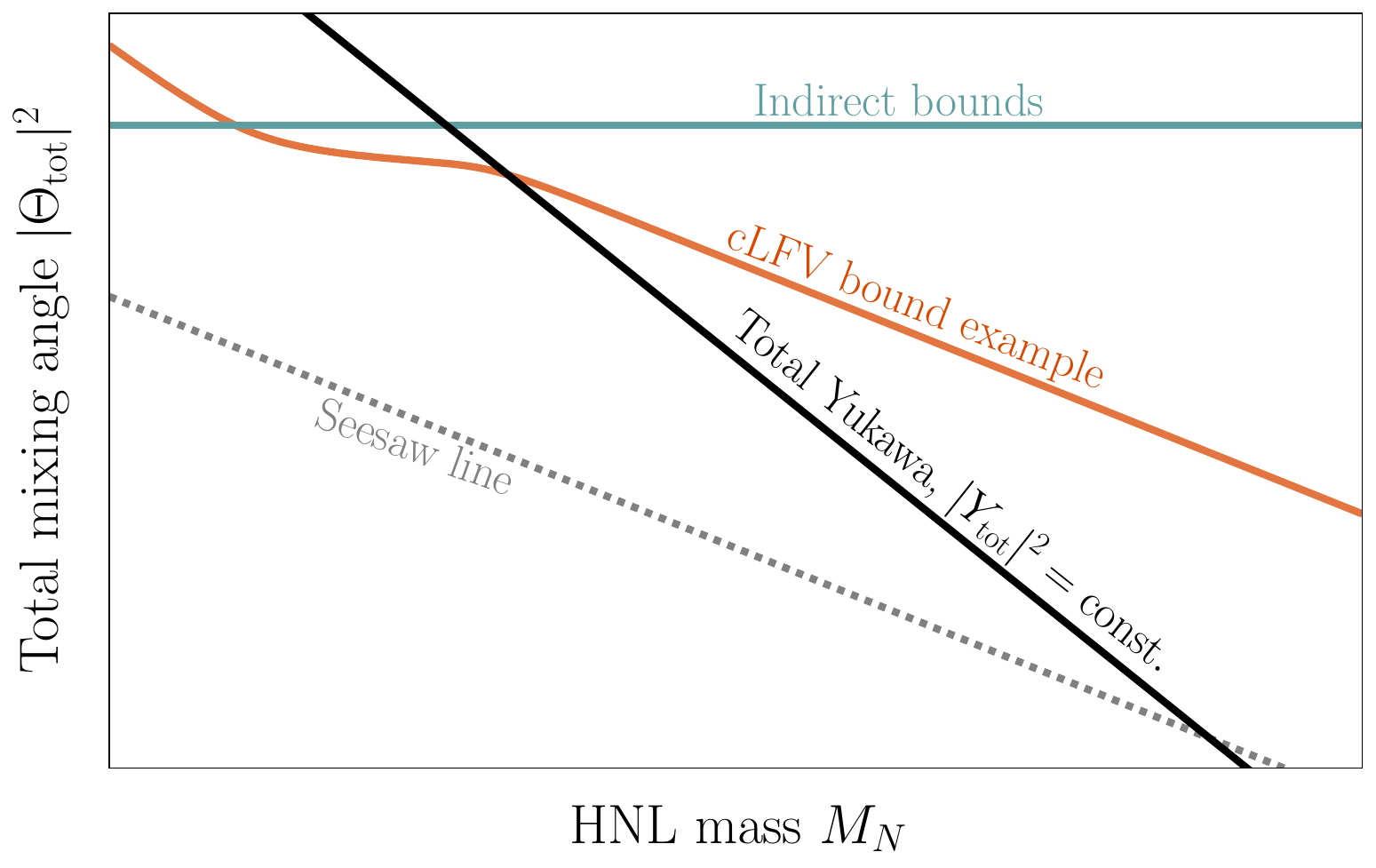}\hfill
    \includegraphics[align = c, width=0.4\linewidth]{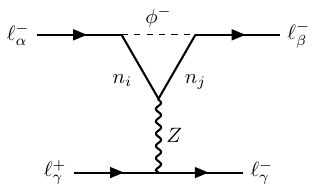}
    \caption{\textit{Left:} Schematic view of the parameter space of the Type-I seesaw model. Below the dotted line, the heavy neutral leptons do not generate neutrino masses at the observed value. The horizontal line (at masses above the electroweak scale) represents a combined fit to electroweak precision data and charged lepton flavor violation searches \cite{Fernandez-Martinez:2016lgt, Blennow:2023mqx}. Corrections to this line appear near the right boundary, where the Yukawa couplings of HNLs are large \cite{Urquia-Calderon:2022ufc}. Determining the exact position of this
    line is the focus of the current paper.
    \textit{Right:} Example of the ``HNL penguin'' diagram, suppressed by the extra powers of the mixing angle $\Theta_\alpha$ but growing with the HNL mass. This diagram dominates rates of several cLFV processes \cite{Urquia-Calderon:2022ufc} and refs.\ therein.    }
    \label{fig:seesaw_parameter_space}
\end{figure}

Collider experiments constrain HNLs with masses below and around the electroweak scale \cite{DELPHI:1996qcc, ATLAS:2015gtp, CMS:2018jxx, CMS:2018iaf,ATLAS:2019kpx,LHCb:2020wxx,ATLAS:2022atq,CMS:2022fut,CMS:2024xdq,CMS:2024ake}. 
Collider bounds quickly deteriorate above the electroweak scale \cite{ATLAS:2023tkz} and one resorts to indirect methods, collectively known as \textit{electroweak precision limits} \cite{Fernandez-Martinez:2016lgt, Blennow:2023mqx}, see Figure~\ref{fig:seesaw_parameter_space}, left panel.
These bounds are dominated by the negative results of searches for charged lepton flavor violation (cLFV) \cite{ParticleDataGroup:2024cfk}. 
Within the type-I seesaw model the cLFV processes are mediated by HNLs running in the loops \cite{Ilakovac:1994kj, Illana:2000ic, Alonso:2012ji, Arganda:2016zvc, Chrzaszcz:2019inj}. 
Among all the loop diagrams there is a notable class of loop diagrams, such as penguin and box diagrams,
that exhibit a so-called ``non-decoupling" behavior \cite{Urquia-Calderon:2022ufc}, i.e.,
the decay width of cLFV processes grows with HNL masses at fixed $\Theta$ \cite{Cheng:1991dy, Tommasini:1995ii, Urquia-Calderon:2022ufc}. 
Figure~\ref{fig:seesaw_parameter_space}, right panel shows one such  diagram for the processes like $\mu\to eee$ and similar conversion process. 
These diagrams provide the dominant contribution to the rate at large masses.
This seemingly counter-intuitive behavior is actually well-known for theories that undergo spontaneous symmetry breaking (see Chapter 8 of \cite{Collins:1984xc} and Refs.~\cite{Collins:1978wz, DHoker:1984izu, DHoker:1984mif, Cheng:1991dy, Tommasini:1995ii, Arganda:2016zvc, Urquia-Calderon:2022ufc}). 
Particularly in the type I seesaw, this ``non-decoupling'' is an artifact of keeping the mixing angle fixed while increasing the HNL mass. 
Such a scaling makes the Yukawa coupling constant grow. 
Instead, if one increases the mass while keeping the Yukawa fixed, heavy HNLs decouple as expected.  
The cLFV bounds determine both the upper bound on the mixing angle for a given HNL mass and the maximal value of the mass, as long as the bounds are consistent with perturbativity. Therefore, we have to know the maximal value of Yukawa constants that is consistent with perturbativity. 
Alternatively, one can understand effect in the language of effective field theory of the Standard Model (SMEFT) \cite{Brivio:2017vri}. There it manifests itself at one-loop matching between the type-I seesaw and SMEFT~\cite{Zhang:2021jdf,Du:2022vso}.
Given quartic dependence of the aforementioned corrections on the Yukawa couplings, the exact position of the perturbativity region becomes of great experimental importance.

We have several tools at our disposal to determine the value of Yukawa couplings which break perturbativity, such as an analysis on the running of the couplings (the so-called \textit{triviality bound}) \cite{Kuti:1987nr}, bounds based on the effects on the running of the Higgs self-coupling to test the vacuum stability \cite{Bambhaniya:2016rbb, Chauhan:2023pur}, and perturbative unitarity (for a recent review, see \cite{Logan:2022uus}).

In this paper, we will specifically use the latter tool, perturbative unitarity, a now-old tool famously used in the 1970s to provide a first ``upper bound'' on the Higgs mass \cite{Dicus:1973gbw, Lee:1977eg, Lee:1977yc}.
The goal of perturbative unitarity is to check for which coupling values does the unitarity of the $S$ matrix be violated for a specific set of processes. It is expected that such an analysis is related to perturbativity, because the inclusion of higher-order corrections will \textit{save} the unitarity of the theory \cite{Logan:2022uus}.
The aforementioned analysis on the Higgs sector was done with tree-level scatterings that are proportional to the Higgs mass itself.

Perturbative unitarity has also been used on the top quark before its  discovery \cite{Chanowitz:1978uj, Chanowitz:1978mv}, as well as other non-minimal Higgs sectors \cite{Horejsi:2005da, Hally:2012pu, Hartling:2014zca}, generic Yukawa and vector interactions \cite{Allwicher:2021rtd, Barducci:2023lqx}, on effective field theory operators \cite{Corbett:2014ora, Corbett:2017qgl}.

Perturbative unitarity has never been applied to the minimal type-I seesaw (outside the same-sign $WW \to \ell \ell$ scattering, considered in \cite{Fuks:2020att}).
In this paper, we will do a systematic analysis to all $2\to 2$ scatterings that all proportional to Yukawa couplings.

The paper is organized as follows: Section~\ref{sec:theoretical_preliminaries} reviews the necessary theoretical background, including both perturbative unitarity and type I seesaw theory. Section~\ref{sec:scattering_amplitudes} summarizes all the processes we considered and presents the results on partial waves and the
constraints on the Yukawa parameters. Section~\ref{sec:bounds_on_bounds} presents how our results affect current bounds, Section~\ref{sec:seesaw_line} presents the same results at the Seesaw line, which gives us results an ``upper bound" on HNL masses. 
Section~\ref{sec:beyond_2HNLs} generalizes our results to models with more HNLs, and discusses limitations to applying these our results to other models with HNLs.
Finally, Section~\ref{sec:conclusions} summarizes and concludes the paper.

\section{Theoretical preliminaries} \label{sec:theoretical_preliminaries}

\subsection{Condition on partial waves} \label{sec:partial_wave_unitarity_bounds}
In this section, we re-derive the well-known condition on partial waves (or the amplitude at a specific total angular momentum, $J$) from the unitarity of the $S$ matrix.

Let us begin by considering $2 \to 2$ processes. The amplitude, $\mathcal{M}_{if}$ (which is defined from the $S$ matrix: $\mel{f}{S - I}{i} = i (2\pi)^4 \delta^4(p_i - p_f) \mathcal{M}_{if}$), for such processes can be decomposed as a series of partial waves \cite{Jacob:1959at}
\begin{align}
    \mathcal{M}_{if}(\theta) = 16\pi \sum_{J} (2J + 1) d_{\mu_i \mu_f}^{J}(\theta) \,a^J_{if}
\end{align}
where $d_{\mu_i, \mu_f}^J(\theta)$ are the Wigner $d$-functions (defined in Appendix \ref{app:def_and_con}), $\mu_i = \lambda_{i1} - \lambda_{i2}$ and $\mu_f = \lambda_{f1} - \lambda_{f2}$ are the difference of the helicity indices of the incoming and outgoing particles respectively, and $a^J_{if}$ are the partial waves.

We can extract the shape of the partial wave if we already have the amplitude
\begin{align}
    a_{if} = \frac{1}{32\pi}\int_{-1}^{1} \dd(\cos\theta) d_{\mu_i \mu_f}^{J}(\theta) \mathcal{M}_{if}(\theta)\,,
\end{align}

The unitarity condition of the $S$ results in the well-known generalized optical theorem
\begin{align} \label{eq:generalized_optical_theorem}
    i\left(\mathcal{M}_{fi}^\ast - \mathcal{M}_{if} \right) = (2\pi)^4 \sum_X \int \dd \Pi_X\,\delta^4(p_i - p_X) \mathcal{M}_{iX}\,\mathcal{M}_{fX}^\ast\,, 
\end{align}
where $\dd \Pi_X = \prod_{j} \frac{\dd^3 p_j}{(2\pi)^3} \frac{1}{2 E_j}$ and $X$ sums over single and multi-particle intermediate states. If we restrict the left-hand side of Eq.~\eqref{eq:generalized_optical_theorem} to $\theta = 0$, 
the right-hand side up to only include two-particle states in $X$ at very high energies 
and set $i = f$, we arrive at:
\begin{align} 
    \label{eq:partial_wave_full_inequality}
    \Im[a_{ii}^J ] \geq \sum_{H} \abs{a_{iH}}^2\,,
\end{align}
where $H$ includes any two-particle state.  
Adding only over $H = i$
we arrive at an inequality that is solved in the complex plane. The solutions are
\begin{align} \label{eq:inequalities_for_partial_waves}
    \abs{a_{ii}} &\leq 1\,, & 0 \leq \Im[a_{ii}] &\leq 1\,, & \abs{\Re[a_{ii}]} &\leq \frac{1}{2}\,.
\end{align}

Partial waves obey this set of inequalities at all orders of perturbation theory for any complete model.
We must comment here that we did the the above analysis assuming that the amplitudes $\mathcal{M}_{ii}$ (and by extension the partial waves, $a_{ii}$) are at all orders of perturbation theory. 
We cannot do this analysis using only the lowest order contribution of perturbation theory because elastic amplitudes at tree level are real at very high energies, and therefore Eq.~\eqref{eq:partial_wave_full_inequality} would be senseless.
However, we will still reserve the right to apply the inequalities in Eq.~\eqref{eq:inequalities_for_partial_waves} to tree-level amplitudes, in particular the third inequality will give us the most constraining results.

In Section \ref{sec:scattering_amplitudes} we will deal with different possible processes (both elastic and inelastic) that can form a system that is expressed as a matrix. The
strongest constraints
will then come from the biggest eigenvalue of said matrix. This will be elucidated better in Section \ref{sec:scattering_amplitudes}.

\subsection{Type I seesaw}
In this section, we will review the necessary components of the type I seesaw. 

Type I seesaw is an extension to the SM that adds to it $\mathcal{N}$ singlet, neutral fermionic fields, $N_{i,R}$. These are usually known as sterile neutrinos, right-handed neutrinos, or heavy neutral leptons (HNLs). The addition of these fields generates new terms for the SM Lagrangian
\begin{align} \label{eq:lagrangian_seesaw}
    \mathcal{L} = \mathcal{L}_{\mathrm{SM}} + i \sum_{i} \bar{N}_{i,R} \slashed{\partial} N_{i,R} - \sum_{\alpha,i}\bar{L}_{\alpha, L} \cdot \tilde{H}\,Y_{\alpha i} N_{i,R} - \frac{1}{2} \sum_{i,j}\bar{N}_{i,R}^C\,(M_M)_{ij}\,N_{j,R} + \mathrm{H.c.}\,,
\end{align}
where $L_{\alpha,L} = \frac{1}{2}(1 - \gamma_5)\,\left( \begin{smallmatrix}
    \nu_{\alpha} \\ \ell_{\alpha}
\end{smallmatrix}\right)$ is the usual SM leptonic doublet, $\tilde{H} = i \sigma_2 H^\ast$ and $H = \left( \begin{smallmatrix}
    \phi^+ \\ \phi^0
\end{smallmatrix}\right)$ is the SM scalar doublet, $Y$ is a Yukawa matrix, $M_M$ is the Majorana mass matrix of $N$, and the sums go from $i = 1, \dots, \mathcal{N}$ and $\alpha = e, \mu, \tau$.

After spontaneous symmetry breaking, the Yukawa term generates an additional mass term that mixes the $\boldsymbol{\nu} = \left(\begin{smallmatrix}
    \nu_e \\ \nu_\mu \\ \nu_\tau
\end{smallmatrix}\right)$ and the $\boldsymbol{N} = \left(\begin{smallmatrix}
    N_1  \\ \vdots \\ N_\mathcal{N}
\end{smallmatrix}\right)$ fields
\begin{align} \label{eq:mass_matrix}
    \mathcal{L}_\mathrm{mass} = -\frac{1}{2} \begin{pmatrix}
        \bar{\boldsymbol{\nu}}_L & \bar{\boldsymbol{N}}_R^C
    \end{pmatrix} 
    \begin{pmatrix}
        0 & M_D \\ M_D^T & M_M
    \end{pmatrix}
    \begin{pmatrix}
        \boldsymbol{\nu}_L^C \\ \boldsymbol{N}_R
    \end{pmatrix} + \mathrm{H.c.}\,,
\end{align}
where $(M_D)_{\alpha i} = \frac{v}{\sqrt{2}} Y_{\alpha i}$ and $v$ is the Higgs vacuum expectation value (vev). We can obtain the mass spectrum of the theory by diagonalizing the matrix in Eq.~\eqref{eq:mass_matrix}. The subsequent diagonalization results in an equation that relates the masses of SM neutrinos, $\boldsymbol{\nu}$, and of our new singlets, $\boldsymbol{N}$. The leading term of this relation is
\begin{align} \label{eq:seesaw_relation}
    V\,m_\nu\,V^T \simeq - \Theta M_N \Theta^T\,, 
\end{align}
where $V$ is the PMNS matrix the matrix that diagonalizes the neutrino mass matrix, $\Theta = M_D\, M_N^{-1}$ is the mixing angle that helps diagonalize the mass matrix and $M_N \simeq M_M$. The dimensions of $M_D$, $\Theta$, and $M_M$ depend on the number of singlets we choose to add, but two is the minimal number such that the model can explain neutrino oscillation experiments. 

\subsubsection{Yukawa term}
The equalities detailed above allow us to relate the Yukawa matrix $Y$ with measurable parameters, like $\Theta$ and $M_N$:
\begin{align} \label{eq:yukawa_definition}
    Y_{\alpha i} \simeq \frac{g}{\sqrt{2} M_W} \Theta_{\alpha i} (M_N)_{ii}\,.
\end{align}  

We are particularly interested in how large the value of the Yukawa matrix can be such that the unitarity of the $S$ matrix is respected. We shall work in the ultra-high-energy ($\sqrt{s} \gg M_N, M_W, M_Z, M_H$) regime where interactions with longitudinal $W$ and $Z$ bosons dominate. 

In this energy limit, according to the Goldstone Equivalence theorem, amplitudes with external longitudinal gauge bosons, $W^\pm_L, Z_L$, are related to amplitudes with external Goldstone bosons, $\phi^\pm, \phi_Z^0$ by \cite{Lee:1977eg, Lee:1977yc, Chanowitz:1985hj, Gounaris:1986cr, Yao:1988aj, Bagger:1989fc, Veltman:1989ud, He:1992nga, He:1993qa, He:1993yd, Denner:1996gb}
\begin{align}
    \mathcal{M}(W_L^\pm, Z_L, \dots) = (i\,C)^n\,\mathcal{M}(\phi^\pm, \phi_Z^0, \dots) + \mathcal{O}\left(M_W/\sqrt{s} \right)\,,
\end{align}
where $n$ is the number of external Goldstone bosons. $C$ is a constant related to the renormalization scheme we choose. Since we will only be working with amplitudes at tree-level, we can set $C = 1$. 

We can thus only consider the interactions that stem from the Yukawa term in Eq.~\eqref{eq:lagrangian_seesaw}.

We can parametrize any Yukawa matrix that follows the seesaw relation in Eq.~\eqref{eq:seesaw_relation} as
\begin{equation}
    Y \simeq i \frac{g}{\sqrt{2} M_W}\, V\,\sqrt{m_\nu}\,O\,\sqrt{M_N}\,,
\end{equation}
where $O$ is an arbitrary orthogonal matrix (or semi-orthogonal, in the case we do not have 3 HNLs). This parametrization is called the \textit{Casas-Ibarra parametrization}~\cite{Casas:2001sr}. We will examine the simplest case, where we only have 2 HNLs with degenerate masses since this is the minimal model that can explain neutrino oscillations. In this case, $O$ can take the shapes
\begin{align} \label{eq:casas_ibarra_matrices}
    O_{\mathrm{NO}} &= \begin{pmatrix}
        0 & 0 \\ 
        \cos\omega & \sin\omega \\ 
        -\sin\omega & \cos\omega
    \end{pmatrix}\,, 
    &O_{\mathrm{IO}} = \begin{pmatrix}
        \cos\omega & \sin\omega \\ 
        -\sin\omega & \cos\omega \\
        0 & 0 
    \end{pmatrix}\,,
\end{align}
where the $\mathrm{NO}$ or $\mathrm{IO}$ subscripts indicate whether neutrinos follow the normal or inverted mass hierarchy, respectively. 

We can simplify even further. If we want values of $\Theta$ that can be probed by current experiments, then we require $\Im(\omega) \gg 1$. Then, we approximate the $O$ matrices as
\begin{align}
    O_{\mathrm{NO}} &\approx \frac{e^{-i\omega}}{2}\begin{pmatrix}
        0 & 0 \\ 
        1 & -i \\ 
        i & 1
    \end{pmatrix}\,, 
    &O_{\mathrm{IO}} \approx \frac{e^{-i\omega}}{2}\begin{pmatrix}
        1 & -i \\ 
        i & 1 \\ 
        0 & 0
    \end{pmatrix}\,.
\end{align}
In this shape, both matrices are rank one. This implies that only one specific linear combination of HNLs interacts with another specific linear combination of the lepton doublet (see also \cite{Shaposhnikov:2006nn, Kersten:2007vk}). These linear combinations are
\begin{equation}
\label{eq:linear_combinations_that_matter}
\begin{aligned}
    N_R &= \frac{1}{\sqrt{2}}\left(N_{1,R} + i N_{2,R} \right)\,, \\
    \nu_L &= \frac{1}{\sqrt{m_{\nu,i} + m_{\nu,j}}} \left(\sqrt{m_{\nu,j}}\,\nu_{j,L} - i \sqrt{m_{\nu,i}}\,\nu_{i,L}  \right)\,,\\
    \ell_L &= \frac{1}{\sqrt{m_{\nu,i} + m_{\nu,j}}} \left(\sqrt{m_{\nu,j}}\,\ell_{j,L} - i \sqrt{m_{\nu,i}}\,\ell_{i,L}  \right)\,,
\end{aligned}
\end{equation}
where $L_{i,L} = \sum_{\alpha}\,V_{\alpha i}\,L_{\alpha,L}$, and $(i,j) = (2,3)$ for normal ordering and $(i,j) = (1,2)$ for inverted ordering, and $m_{\nu,1}, m_{\nu,2}$ and $m_{\nu,3}$ are the masses of light neutrinos from neutirno oscillation data. Their values are \cite{Esteban:2020cvm, deSalas:2020pgw}
\begin{align}
    m_{\nu,2} &= \SI{8.61e-3}{eV}\,, & m_{\nu,3} &= \SI{5.01e-2}{eV} & &\text{for Normal Ordering}\,, \\
    m_{\nu,1} &= \SI{4.72e-2}{eV}\,, & m_{\nu,2} &= \SI{4.79e-2}{eV} & &\text{for Inverted Ordering}\,,
\end{align}

In this scenario, the interaction term in Eq.~\eqref{eq:lagrangian_seesaw} becomes
\begin{equation}
    \label{eq:yukawa_interaction}
    \begin{aligned}
        \mathcal{L}_Y &= -Y_{\mathrm{tot}}\,\bar{L}_L \cdot \tilde{H}\,N_R + \mathrm{H.c.}\,, \\[2ex]
        &= -Y_{\mathrm{tot}} \left[\bar{\nu}_L\,N_R\,\phi^{0\ast} - \bar{\ell}_L\,N_R\,\phi^- \right] + \mathrm{H.c.}\,,
    \end{aligned}
\end{equation}
where $Y_{\mathrm{tot}} = \frac{g}{2 M_W} e^{-i \omega} \sqrt{M_N\,(m_{\nu,i} + m_{\nu,j})}$. Notice that
\begin{align}
    \abs{Y_{\mathrm{tot}}}^2 = \sum_{\alpha,i} \abs{Y_{\alpha i}}^2 \simeq \frac{g^2\,M_N\,(m_{\nu,i} + m_{\nu,j})}{4 M_W^2}\,e^{2\Im(\omega)}\,,
\end{align}
and $\phi^0 = h + i \phi_Z^0$ where $h$ is the Higgs field and $\phi_Z^0$ is the Goldstone boson that is eaten by the $Z$ boson.

The reason why we are expressing our Lagrangian in terms of $\phi^0$ and not the usual scalar fields ($h$ and $\phi^0_Z$) is because $\phi^0$ has a defined weak isospin and hypercharge. We can recognize the $\mathrm{SU}(2) \times \mathrm{U}(1)$  structure that the interactions have when working in this basis.

In our previous derivation, it might seem intriguing that a rank one matrix can seemingly explain neutrino oscillation data, since intuitively it could only provide mass to one neutrino. 
Moreover, due to having degenerate HNL masses, there is a lepton symmetry that leaves this neutrino massless. 
However, let us consider again the full matrix in Eq.~\eqref{eq:casas_ibarra_matrices}, as a sum of two rank one matrices
\begin{align}
    O_\mathrm{NO} = \frac{e^{-i\omega}}{2} \begin{pmatrix}
        0 & 0 \\
        1 & -i \\
        i & 1
    \end{pmatrix} + \frac{e^{i\omega}}{2} \begin{pmatrix}
        0 & 0 \\
        1 & i \\
        -i & 1
    \end{pmatrix}\,, \\[2ex]
    O_\mathrm{IO} = \frac{e^{-i\omega}}{2} \begin{pmatrix}
        1 & -i \\
        i & 1 \\
        0 & 0 
    \end{pmatrix} + \frac{e^{i\omega}}{2} \begin{pmatrix}
        1 & i \\
        -i & 1 \\
        0 & 0
    \end{pmatrix}\,,
\end{align}
and first rank one matrices on both equalities is the one we considered in our analysis above. 
Considering only one of these two matrices alone will not generate the neutrino masses necessary to explain oscillation, however considering the both of them will reproduce it.

However, for the interactions we will consider below, only the part of the Casas-Ibarra matrix proportional to $e^{-i\omega}$ will be relevant, since it will be proportional to $e^{2 \Im\omega}$, as we showed above.
The part proportional to $e^{i\omega}$ will be proportional to $e^{-2 \Im\omega}$, which will be completely suppressed in the limit $\Im(\omega) \gg 1$. Any interference between both matrices, will also be suppressed. 
But, when $\Im(\omega) \simeq 0$, it will be relevant. This is close to the so-called \textit{seesaw line}, and will be the main focus of Section~\ref{sec:seesaw_line}.

It is straightforward to compute the Feynman Rules of the interactions in Eq.~\eqref{eq:yukawa_interaction}. The only caveat comes from the Majorana nature of neutrinos and HNLs. The Majorana nature of both fields gives rise to different Feynman rules from those of usual Dirac particles, but the difference between both is proportional to the masses of HNLs and neutrinos. For our analysis, we can ignore this difference and treat them as Dirac particles, this is the core of the so-called \textit{Majorana-Dirac confusion theorem} \cite{Kayser:1981nw,Kayser:1982br, Kayser:1984xc, Zralek:1997sa}.

Some processes we are about to consider receive contributions proportional to the SM gauge couplings and not just our new Yukawa parameters. We neglect them since we are working on the regime where $\abs{Y_{\mathrm{tot}}}^2 \sim \mathcal{O}(1)$, which is much bigger than other SM parameters.

Our results in Section~\ref{sec:scattering_amplitudes} will work for any parametrization of Yukawa matrices with rank one. 
This is the case for models with both two and three degenerate HNLs, with a large mixing angle. 
We will elaborate on this comment in Section~\ref{sec:beyond_2HNLs}.

\section{Scattering amplitudes and partial waves} \label{sec:scattering_amplitudes}
We computed all the amplitudes and partial waves with the help of Mathematica packages 
\texttt{FeynCalc 9.3.1} \cite{Shtabovenko:2016sxi} and \texttt{FeynArts 3.11} \cite{Hahn:2000kx}. We obtained the FeynArt amplitudes from a custom-made \texttt{FeynRules} \cite{Alloul:2013bka} file specifically used for our purposes, which includes the Yukawa interactions we are interested in.

For our analysis, we focus on scattering processes that involve only scalars and fermions. These scatterings are proportional to $\abs{Y_\mathrm{tot}}^2$ and correspond to angular momenta of $J = 0, \frac{1}{2}, 1$. The minimal type I seesaw also allows for higher values of $J$ due to interactions between HNLs and the transverse parts of the gauge bosons which we neglect, as they are proportional to the weak coupling constants and the HNL mixing angles.

The angular momentum of each process depends on the initial and final helicities of the particles in the model. We summarize all the processes we will analyze in Table~\ref{tab:possible_amplitudes}. There is also the possibility of having lepton number violating processes (LNV) due to the Majorana nature of HNLs and neutrinos, such as $\ell^- \phi^+ \to \ell^+ \phi^-$, but all such  processes are suppressed by $M_N^2/s$ (an analysis on such processes was previously done in \cite{Fuks:2020att}) and are negligibly small for the ultra-high-energy processes we are considering.

\begin{table}[t]
        \centering
        \hspace*{-1.75cm}
        \begin{tabular}{c c  c c c  c c c c }
            \cline{3-9} & & \multicolumn{3}{c}{$\mu_i = 0$} & \multicolumn{2}{c}{
            $\mu_i = \pm \frac{1}{2}$} & \multicolumn{2}{c}{$\mu_i = \pm 1$} \\
            & & $++$ & $--$ & $00$ & $+0$ & $-0$ & $+-$ & $-+$  \\ \hline 
            \multirow{3}{3.825em}{$\mu_f = 0$} 
            & $++$ & \cellcolor{blue!10}$++ \to ++$ & \cellcolor{blue!10} & \cellcolor{blue!10} & & & \cellcolor{red!15} & \cellcolor{red!15} \\
            & $--$ & \cellcolor{blue!10} & \cellcolor{blue!10}$-- \to --$ & \cellcolor{blue!10} & & & \cellcolor{red!15} & \cellcolor{red!15} \\ 
            & $00$ & \cellcolor{blue!10} & \cellcolor{blue!10} & \cellcolor{blue!10} & & & \cellcolor{red!15} $+- \to 00$ & \cellcolor{red!15} $- + \to 00$  \\ \hline
            \multirow{2}{3.825em}{$\mu_f = \pm \frac{1}{2}$} 
            & $+0$ & & & & \cellcolor{yellow!15} $+ 0 \to + 0$ & \cellcolor{yellow!15} & &  \\
            & $-0$ & & & & \cellcolor{yellow!15} & \cellcolor{yellow!15} $-0 \to - 0$ & & \\ \hline
            \multirow{2}{3.825em}{$\mu_f = \pm 1$} 
            & $+ -$ &  \cellcolor{red!15} & \cellcolor{red!15} & \cellcolor{red!15} $00 \to +-$ & & & \cellcolor{red!15} $+- \to +-$ & \cellcolor{red!15} $-+ \to +-$ \\
            &$- +$ & \cellcolor{red!15} & \cellcolor{red!15} & \cellcolor{red!15} $00 \to -+$ & & & \cellcolor{red!15} $+- \to -+$ & \cellcolor{red!15} $-+ \to -+$ \\
            \hline
        \end{tabular}
        \caption{Table of all possible scatterings with their respective values of angular momentum that can be mediated by $\abs{Y_\mathrm{tot}}^2$. It should be noted that the $+$ and $-$ notation is an abbreviation for $+\frac{1}{2}$ and $-\frac{1}{2}$. The processes with the blue background are the processes with $J = 0$ angular momentum transferred, the ones in yellow are with $J = \frac{1}{2}$, and the red ones with $J = 1$.}
        \label{tab:possible_amplitudes}
\end{table}

\subsection{\texorpdfstring{$J = 0$}{Lg} amplitudes}
The processes with $J=0$ partial waves are the scatterings of fermions where both initial and final fermions have the same helicities. These processes are
\begin{align}
    \label{eq:j_0_process_1}
    N_\pm\,\ell_\pm^\pm &\leftrightarrow N_\pm\,\ell_\pm^\pm\,, \\
    \label{eq:j_0_process_2}
    N_\pm\,\nu_\pm &\leftrightarrow N_\pm\,\nu_\pm\,, 
\end{align}
where the sub-indices denote the helicity of the particle. Of course, at very high energies, we would expect the negatively charged leptons to have negative helicity due to the chiral structure of Yukawa interactions (see Eq.~\eqref{eq:yukawa_interaction}). For neutrinos, since they are Majorana particles, we expect both positive and negative helicities to be able to interact. In the literature, this is usually shown with a \textit{symmetrization} in the $\nu-N-\phi^0$ term in Eq.~\eqref{eq:yukawa_interaction}, which we did not show in this instance. 

These two scattering processes are due to an $s$ and a $u$ channel. Specifically, the $++ \to ++$ and the $-- \to --$ processes have angular momentum $J = 0$, which are due to an $s$ channel. Both channels provides the same
constraint
on the Yukawa couplings because they all have the same partial wave
\begin{align}
    a^{J = 0} = -\frac{\abs{Y_{\mathrm{tot}}}^2}{32 \pi}\int_{-1}^1 \dd(\cos\theta) d_{00}^0(\theta)= -\frac{Y_{\mathrm{tot}}^2}{16\pi}\,,
\end{align}
which the unitarity condition on $S$ gives us
\begin{align}
    \label{eq:ytot_8pi_bound}
    \abs{Y_{\mathrm{tot}}}^2 \leq 8\pi\,.
\end{align}

What is surprising is that this
result holds independently of the shape of the Yukawa matrix. We present the proof in Appendix~\ref{app:general_matrices}.

\subsection{\texorpdfstring{$J = \frac{1}{2}$}{Lg} amplitudes}
Scatterings of fermions and scalars are the only processes with partial waves $J = 1/2$. In contrast to the $J=0$ scatterings, here we have more possible processes that can be inelastic or elastic. For example, we have a set of elastic scatterings that involve an HNL in the final and initial states
\begin{align}
    \label{eq:j_1/2_process_1}
    N_\pm\,\phi_0^+ &\leftrightarrow N_\pm\,\phi_0^+\,, \\
    \label{eq:j_1/2_process_2}
    N_\pm\,\phi_0^0 &\leftrightarrow N_\pm\,\phi_0^0\,,
\end{align}
and their respective charge-conjugated processes. Both of these scatterings give the same
restriction on the Yukawa coupling. These processes are due to an $s$ and a $u$ channel, but these two do not to add one another, as they mediate the process with $N_+$ or $N_-$ in Eqs.~\eqref{eq:j_1/2_process_1} and \eqref{eq:j_1/2_process_2} but not both. The
most stringent results
always come from the $u$ channel
\begin{align}
    \label{eq:partial_wave_J=1/2_1}
    &a^{J=1/2}_{NH} = -\frac{\abs{Y_{\mathrm{tot}}}^2}{32 \pi}\int_{-1}^1 \dd(\cos\theta) d_{\frac{1}{2} \frac{1}{2}}^\frac{1}{2}(\theta) \sec(\frac{\theta}{2})= -\frac{Y_{\mathrm{tot}}^2}{16\pi}\,, \\[3ex]
    &\implies \abs{Y_{\mathrm{tot}}}^2 \leq 8\pi\,,
\end{align}
the same
result as for $J=0$ scatterings.

We also have scatterings mediated by HNLs that can be elastic or inelastic. We separate two sets of possible scatterings
\begin{align}
    \label{eq:j_1/2_process_3}
    \left\lbrace\begin{aligned}
        \ell^-_-\,\phi_0^+ &\leftrightarrow \ell_-^-\,\phi_0^+\,, \\
        \ell_-^-\,\phi_0^+ &\leftrightarrow \nu_-\,\phi_0^0\,, \\
        \nu_-\,\phi_0^0 &\leftrightarrow \nu_-\,\phi_0^0\,,
    \end{aligned}\right. \\[2ex]
    \label{eq:j_1/2_process_4}
    \left\lbrace\begin{aligned}
        \nu_-\,\phi_0^- \leftrightarrow \ell_-^-\,\phi_0^{0\ast}\,,
    \end{aligned}\right.
\end{align}

The processes presented in Eqs.~\eqref{eq:j_1/2_process_1} and \eqref{eq:j_1/2_process_2} are both elastic processes, and thus the
inequality in Eq.~\eqref{eq:inequalities_for_partial_waves} apply. But the processes highlighted in Eqs.~\eqref{eq:j_1/2_process_3} contain both inelastic and elastic processes, and the one in \eqref{eq:j_1/2_process_4} is completely inelastic, so naively it is not entirely clear how to deal with them. We can extract the necessary information by writing these processes in a matrix. Let's consider the scatterings in Eq.~\eqref{eq:j_1/2_process_3}, in the basis $\ell_-^-\,\phi_0^+, \nu_-\,\phi_0^0$ we have the following partial wave matrix
\begin{align}
    \label{eq:j_1/2_partial_waves_matrix_1}
    \mathbf{a}^{J = 1/2}_{LH} = \frac{Y_{\mathrm{tot}}^2}{32\pi}\int_{-1}^1\dd(\cos\theta) \,\cos(\frac{\theta}{2})\,d_{\frac{1}{2},\frac{1}{2}}^{\frac{1}{2}}(\theta) \begin{pmatrix}
        - 1 & 1 \\ 1 & -1
    \end{pmatrix} = \frac{\abs{Y_{\mathrm{tot}}}^2}{32\pi} \begin{pmatrix}
        -1 & 1 \\ 1 & -1
    \end{pmatrix}\,,
\end{align}
the inequalities in Eq.~\eqref{eq:inequalities_for_partial_waves} apply to the elements in the diagonal of this matrix, but they also apply to its eigenvalues, and in particular, the strongest
constraint
comes from the biggest eigenvalue. Our labor is then to diagonalize this matrix
\begin{align}
    \label{eq:j_1/2_partial_waves_matrix_2}
    \mathbf{a}^{J = 1/2}_{LH} =\frac{\abs{Y_{\mathrm{tot}}}^2}{32\pi} \begin{pmatrix}
        -1 & 1 \\ 1 & -1
    \end{pmatrix} \to 
    \frac{\abs{Y_{\mathrm{tot}}}^2}{32\pi} \begin{pmatrix}
        -2 & 0 \\ 0 & 0
    \end{pmatrix}\,,
\end{align}
this eigenvalue corresponds to the scattering $\frac{1}{\sqrt{2}}\left[\ell_-^-\,\phi_0^+\, - \,\nu_-\,\phi_0^0\right] \leftrightarrow \frac{1}{\sqrt{2}}\left[\ell_-^-\,\phi_0^+\, - \,\nu_-\,\phi_0^0\right]$. 
These scatterings provide the same constraint
as in Eq.~\eqref{eq:ytot_8pi_bound}.

The same idea applies to the processes in Eq.~\eqref{eq:j_1/2_process_4}. In the $\nu_-\phi_0^-, \ell^-_- \phi_0^{0\ast}$ basis we get the matrix
\begin{equation}
        \mathbf{a}^{J = 1/2}_{L\tilde{H}}  = \frac{\abs{Y_{\mathrm{tot}}}^2}{16\pi}
    \begin{pmatrix}
        0 & 1 \\ 1 & 0
    \end{pmatrix}
    \rightarrow \frac{\abs{Y_{\mathrm{tot}}}^2}{16\pi} \begin{pmatrix}
        1 & 0 \\ 0 & -1
    \end{pmatrix}\,,
\end{equation}
which also leads to the
constraint
$\abs{Y_{\mathrm{tot}}}^2 \leq 8\pi$.

\subsection{\texorpdfstring{$J = 1$}{Lg} amplitudes}
Similar to the $J = \frac{1}{2}$, we have different channels, both elastic and inelastic can be written in a matrix like the one in Eq.~\eqref{eq:j_1/2_partial_waves_matrix_1}. The processes that contribute to $J = 1$ include fermion scatterings $\pm \mp \leftrightarrow \pm \mp$ and $\pm \mp \leftrightarrow \mp \pm$, and scalar-fermion scatterings $\pm \mp \leftrightarrow 0 0$. This includes the scatterings we considered for $J = 0$ in Eqs.~\eqref{eq:j_0_process_1} and \eqref{eq:j_0_process_2}, with the difference that the processes have different helicities
\begin{align}
    \label{eq:j_1_process_1}
    N_{\pm}\,\ell^{\mp}_{\mp} \leftrightarrow N_{\pm}\,\ell^{\mp}_{\mp}\,, \\
    \label{eq:j_1_process_2}
    N_{\pm}\,\nu_{\mp} \leftrightarrow N_{\pm}\,\nu_{\mp}\,, 
\end{align}
and also the following set of scatterings
\begin{align}
    \label{eq:j_1_process_3}
    \ell^-_-\,\nu_+ &\leftrightarrow \phi^-_0\,\phi^0_0\,, \\
    \label{eq:j_1_process_4}
    \left\lbrace N_-\,N_+\,, \nu_-\,\nu_+\,, \ell^-_-\,\ell^+_+\,, \phi^0_0\,\phi^{0\ast}_0\,, \phi^+_0\,\phi^-_0 \right\rbrace &\leftrightarrow \left\lbrace N_-\,N_+\,, \nu_-\,\nu_+\,, \ell^-_-\,\ell^+_+\,, \phi^0_0\,\phi^{0\ast}_0\,, \phi^+_0\,\phi^-_0 \right\rbrace\,.
\end{align}
in Eq.~\eqref{eq:j_1_process_4}, every initial state can go into every final state and vice versa. 

Both scatterings in Eqs.~\eqref{eq:j_1_process_1} and \eqref{eq:j_1_process_2} give the same partial waves. This fact should not be surprising due to the fact that there is an $\mathrm{SU}(2)$ rotation that relates both amplitudes. The results for both are
\begin{align}
    &a^{J = 1}_{NL} = \frac{\abs{Y_{\mathrm{tot}}}^2}{32 \pi} \int_{-1}^1 \dd(\cos\theta) d_{11}^1(\theta) = \frac{\abs{Y_{\mathrm{tot}}}^2}{32\pi}\,, \\[3ex]
    &\implies \abs{Y_{\mathrm{tot}}}^2 \leq 16\pi\,,
\end{align}
which is a weaker result than the ones obtained in previous sections. 

For the scatterings in Eqs.~\eqref{eq:j_1_process_3} in the $\ell_-^-\,\nu_+\,,\phi_0^-\,\phi_0^0$ basis we have
\begin{alignat}{3}
    \mathbf{a}^{J=1}_{LH}
    &= \frac{\abs{Y_{\mathrm{tot}}}^2}{16\sqrt{2}\pi} \begin{pmatrix}
        0 & 1 \\ 1 & 0
    \end{pmatrix} \rightarrow -\frac{\abs{Y_{\mathrm{tot}}}^2}{16\sqrt{2}\pi}  \begin{pmatrix}
        1 & 0 \\ 0 & -1
    \end{pmatrix}\,, \\[3ex]
    \implies \abs{Y_{\mathrm{tot}}}^2 \leq 8\sqrt{2}\pi\,, \hspace*{-2.5cm}
\end{alignat}
which is also weaker than every other result by a factor of $\sqrt{2}$.

Finally, the partial wave matrix in Eq.~\eqref{eq:j_1_process_4} in the $N_-\,N_+\,, \nu_-\,\nu_+\,, \ell^-_-\,\ell^+_+\,, \phi^0_0\,\phi^{0\ast}_0\,, \phi^+_0\,\phi^-_0$ basis reads as
\begin{align}
    \mathbf{a}^{J=1}_{NLH}
    \label{eq:partial_wave_matrix_J1}
    &= \frac{\abs{Y_{\mathrm{tot}}}^2}{32\pi} \begin{pmatrix}
        0 & 1 & 1 & -\sqrt{2} & -\sqrt{2} \\
        1 & 0 & 0 & -\sqrt{2} & 0 \\
        1 & 0 & 0 & 0 & -\sqrt{2} \\
        -\sqrt{2} & -\sqrt{2} & 0 & 0 & 0 \\
        -\sqrt{2} & 0 & -\sqrt{2} & 0 & 0 \\
    \end{pmatrix} \rightarrow \frac{\abs{Y_{\mathrm{tot}}}^2}{32\pi} \begin{pmatrix}
        1 + \sqrt{5} & & & & \\
        & 1 - \sqrt{5} & & & \\
        & & \sqrt{2} & & \\
        & & & -\sqrt{2} & \\
        & & & & -2 
    \end{pmatrix}\,,
\end{align}
\begin{align}
    \label{eq:bound_J=1}
    \implies \abs{Y_{\mathrm{tot}}}^2 \leq \frac{16\pi}{1 + \sqrt{5}} = \frac{8\pi}{\varphi}\,,
\end{align}
where $\varphi = \frac{1 + \sqrt{5}}{2} \simeq 1.618\dots$, is the golden ratio. This is the 
strongest constraint
we could find. The 
constraint would correspond to the scattering 
\begin{align*}
    \frac{1}{2} \sqrt{\frac{\varphi}{\varphi - \frac{1}{2}}}&\left[ N_- N_+ + \frac{1}{\varphi}\left(\nu_- \nu_+ + \ell^-_- \ell^+_+ \right) + \frac{1}{\sqrt{2}}\left(\phi^+ \phi^- + \phi^0 \phi^{0\ast} \right)\right] \\
    &\leftrightarrow \frac{1}{2} \sqrt{\frac{\varphi}{\varphi - \frac{1}{2}}}\left[ N_- N_+ + \frac{1}{\varphi}\left(\nu_- \nu_+ + \ell^-_- \ell^+_+ \right) + \frac{1}{\sqrt{2}}\left(\phi^+ \phi^- + \phi^0 \phi^{0\ast} \right)\right]\,.
\end{align*}

\section{Implications for cLFV and electroweak bounds}
\label{sec:bounds_on_bounds}
\begin{figure}[!t]
    \centering
    \includegraphics[width = \linewidth]{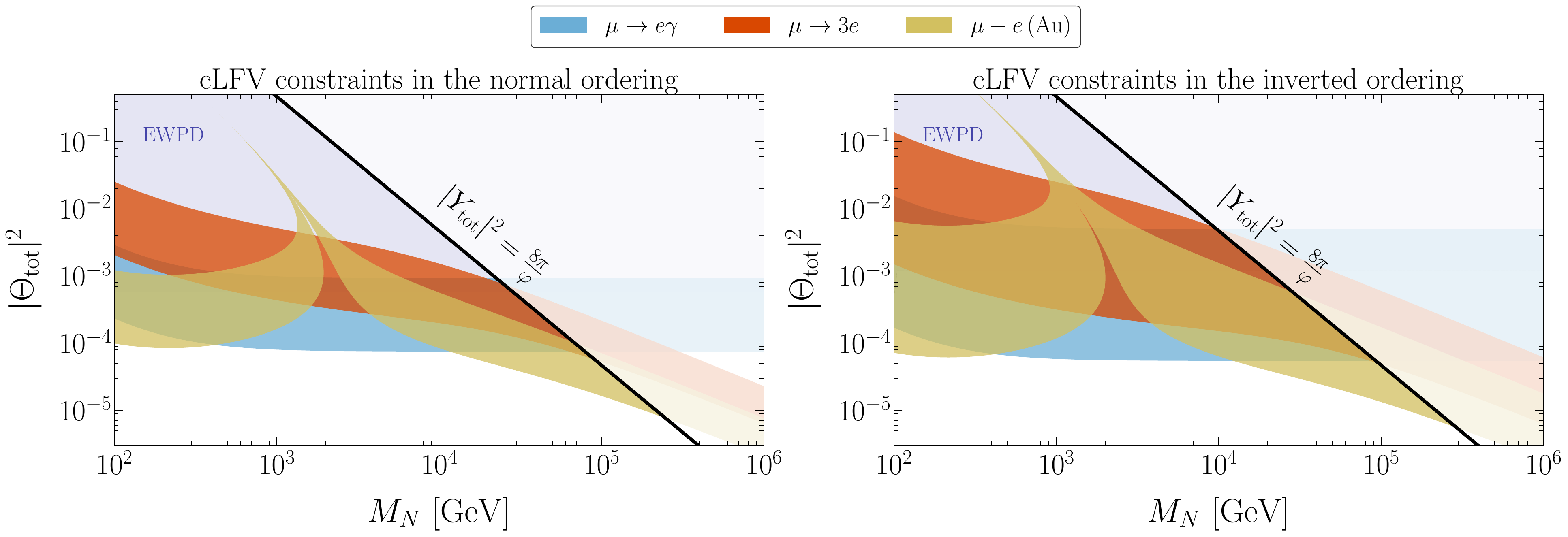}
    \caption{Current bounds on the minimal type-I seesaw with two HNLs. 
    The colorful bands indicate the bounds from cLFV processes, as allowed by neutrino oscillation data (these values are bounded by a triangular defined by a specific ratio of values of the $\Theta$, see, for example, Section 2 of \cite{Tastet:2021vwp}) in the normal (left) and inverted ordering (right). The dark blue shaded region fills the area bounded by EWPD, $|\Theta_{\mathrm{tot}}|^2 < \num{5.8e-4}$ for NO and $|\Theta_{\mathrm{tot}}|^2 < \num{1.2e-3}$ for IO \cite{Blennow:2023mqx}. The black line indicates the line where the unitarity of the $S$ matrix breaks down.}
    \label{fig:clfv_perturbativity}
\end{figure}
From the relation between the Yukawa coupling, HNL masses, and the $\nu-N$ mixing we have in Eq.~\eqref{eq:yukawa_definition}, 
then
if we restrict HNL parameters to hold unitarity for the $2\to 2$ processes we considered, we have
\begin{align}
    \abs{Y_{\mathrm{tot}}}^2 = \frac{g^2\,M_N^2\,\abs{\Theta_{\mathrm{tot}}}^2}{2\,M_W^2} = 2\sqrt{2}\,G_F\,\abs{\Theta_{\mathrm{tot}}}^2 M_N^2  &\leq \frac{8\pi}{\varphi}\,, \\[2ex]
    \implies \Theta_{\mathrm{tot}}^2 M_N^2 &\lesssim (\SI{686.28}{GeV})^2 \label{eq:unitarity_mixing_mass}
\end{align}
where $\Theta_{\mathrm{tot}}^2 = \sum_{\alpha,i} \abs{\Theta_{\alpha i}}^2$. In terms of Casas-Ibarra parameters, it would translate to
\begin{align}
    \frac{1}{2}\,M_N \left(\textstyle\sum_i m_{\nu,i}\right) \exp(\Im\omega) \lesssim (\SI{686.28}{GeV})^2\,.
\end{align}
We remind the reader that these results are only valid for $\mathrm{Im}\omega \gg 1$, way above the seesaw line, the only region of the parameter space that current experiments can probe.

Indirect constraints from the electroweak precision observables place bounds on $\abs{\Theta_{\alpha i}}^2$, and we expect them to be smaller than \num{e-3} (see \cite{Fernandez-Martinez:2016lgt, Chrzaszcz:2019inj, Blennow:2023mqx}). Current bounds place $\abs{\Theta_{\mathrm{tot}}}^2 < \num{5.8e-4}$ for the Normal ordering, and $\abs{\Theta_{\mathrm{tot}}}^2 < \num{1.2e-3}$ for the Inverted ordering \cite{Blennow:2023mqx}, for the minimal model with two HNLs. 
From these bounds and from Eq.~\eqref{eq:unitarity_mixing_mass} we can therefore say that the masses of HNLs cannot be larger than $~\SI{28.5}{TeV}$ for Normal Ordering, and than $~\SI{19.81}{TeV}$ for Inverted Ordering without violating tree-level unitarity for the set of $2\to 2$ processes we considered, unless the mixing angles are smaller.

Current direct probes of HNL models cannot reach such high values of mass and low values of mixing angle, however other indirect probes such as bounds from searches of charged lepton flavor violating processes (cLFV) can reach these ranges.
The most stringent of these processes includes decays such as $\mu\to e\gamma, \mu\to 3e$, and muon conversion in a nucleus. 
HNLs can mediate such processes due to having them running inside of loop diagrams \cite{Ilakovac:1994kj, Illana:2000ic, Alonso:2012ji, Chrzaszcz:2019inj, Granelli:2022eru}.
In particular $\mu \to 3e$, and muon conversion in a nucleus exhibit a behavior called \textit{non-decoupling}, where the amplitude of these processes increases in mass of HNLs \cite{Cheng:1991dy, Tommasini:1995ii, Urquia-Calderon:2022ufc}
for fixed values of $\Theta_{\alpha,i}$\footnote{This is not unique to HNLs, we would expect this for any theory that undergoes spontaneous symmetry breaking (see Chapter 8 of \cite{Collins:1984xc})}.
However, this behavior disappears once we parametrize the amplitudes in terms of elements of Yukawa matrices instead of mixing angles.

All cLFV processes depend heavily on individual flavor mixings. In the case of cLFV processes involving muons and electrons, the dependence looks as $\Gamma \propto \sum_{i} \abs{\Theta_{\mu,i}^{\phantom{\ast}}\,\Theta_{e,i}^\ast}^2$, which can easily be related to $\abs{\Theta}^2$, but still depends on different ratios of mixing between electron and muons.
The ratios are not completely free, and are restricted by neutrino oscillation data (see, for example, Section 2 of \cite{Tastet:2021vwp} or \cite{Drewes:2022akb}). 
We plot the bands of the most optimistic and conservative bounds from cLFV bounds in Fig~\ref{fig:clfv_perturbativity}.

We can see that, in the most stringest of cases, we cannot have HNLs heavier than $\sim \SI{240}{TeV}\,(\sim\SI{280}{TeV})$ for Normal Ordering (Inverted Ordering) from all cLFV processes, without breaking unitarity for the $2\to 2$ processes we analyzed, unless we have smaller mixing angles than \num{1e-5}.

\section{Where unitarity meets seesaw: maximal HNL mass in type-I seesaw model}
\label{sec:seesaw_line}
As we mentioned at the end of Section~\ref{sec:theoretical_preliminaries}, all the results in Section~\ref{sec:scattering_amplitudes} are only valid for large values of $\Im(\omega)$, where the Yukawa matrix is approximately a rank one matrix. 
For $\omega \simeq 0$, the Yukawa matrix becomes a rank two matrix, and we can no longer treat the theory as if one HNL interacts only with one lepton doublet. Instead we have two HNL generations interacting with two generations of the lepton doublet. 

The general shape of the partial wave matrices with three different flavors of leptons and $\mathcal{N}$ different HNL generations is in Appendix~\ref{app:general_yukawa}. In general, obtaining the eigenvalues in the most general case is not possible to do analytically. We only present some analytic results for the $2 + 2$ case, which covers all the parameter space of the minimal type I seesaw with only two HNLs.

\begin{figure}[!t]
    \centering
    \includegraphics[width = \linewidth]{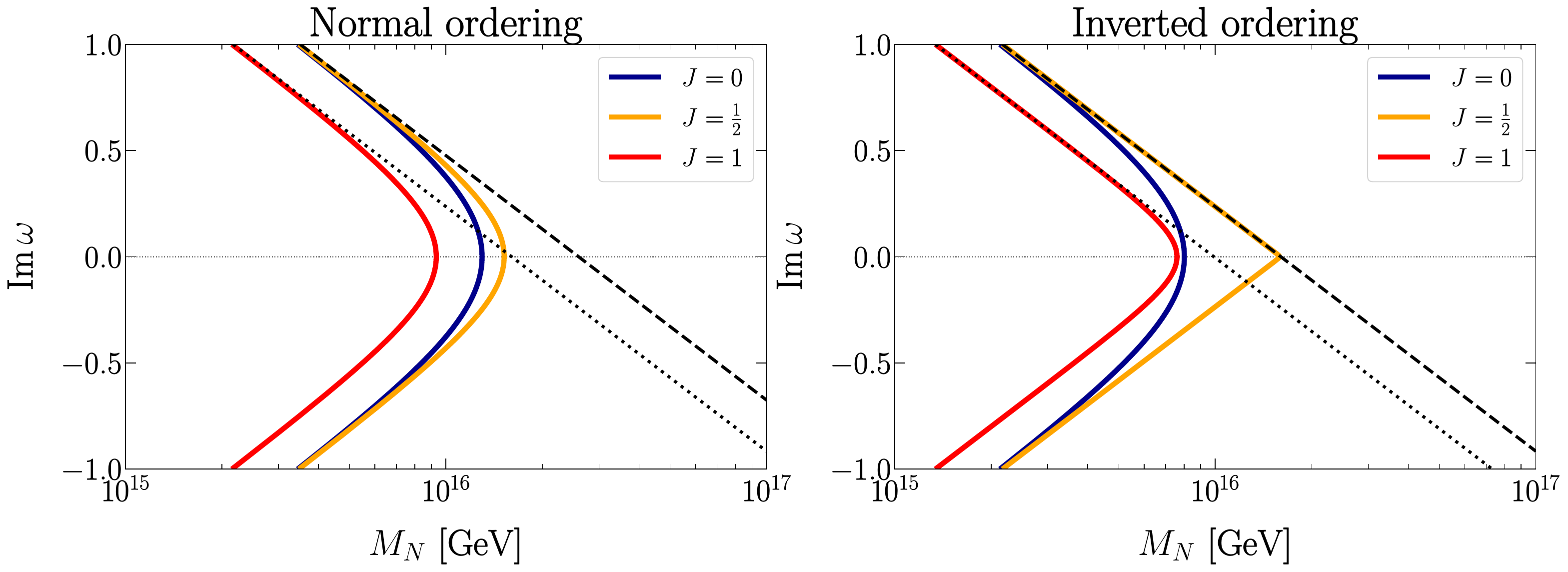}
    \caption{Comparison with the different allowed values of the parameter space, such that tree-level perturbativity is allowed for different values of $\Im(\omega)$ and $M_N$. The black dashed, and dotted lines are the results obtained in Section~\ref{sec:scattering_amplitudes}, whereas the colored lines would be the more exact solution for arbitrary values of $\Im(\omega)$ (read the main text for more information). Everything to the left of the lines is allowed by tree-level unitarity.}
    \label{fig:seesaw_line_unitarity}
\end{figure}

The easiest case is for $J = 0$, where the results of Section~\ref{sec:scattering_amplitudes} are still valid because the general shape of the partial wave matrix is itself a rank one matrix. We explicitly show this fact at the start of Appendix~\ref{app:general_matrices}. 
The change comes when writing the
inequalities in terms of Casas-Ibarra parameters
\begin{align}
    \label{eq:bound_general_casasIbarra_J=0}
    \abs{Y_\mathrm{tot}}^2  = \frac{g^2}{2 M_W^2}\,M_N\,\left(\textstyle\sum_i m_{\nu,i}\right) \cosh(2\Im\omega) \leq 8\pi\,,
\end{align}

We can also do the $J = \frac{1}{2}$ cases analytically. 
From Appendix~\ref{app:general_matrices}, the general for of the partial waves matrices in Eqs.~\ref{eq:j_1/2_partial_waves_matrix_1} and \ref{eq:j_1/2_partial_waves_matrix_2} are
\begin{align} 
    \begin{aligned}
        \mathbf{a}^{J=\frac{1}{2}}_{NH} &= -\frac{1}{16\pi}\,Y\,Y^\dagger\,,  \\[2ex]
        \mathbf{a}^{J=\frac{1}{2}}_{L\tilde{H}} &= -\frac{1}{16\pi}\,Y^\dagger\,Y\,,
    \end{aligned}
\end{align}
where $Y$ is the Yukawa matrix. For our case, this translates into obtaining the eigenvalues of a $2\times 2$ matrix, which gives us
\begin{align}
    \label{eq:bound_general_casasIbarra_J=1/2}
    \frac{g^2}{4 M_W^2} M_N \left[\left(m_{\nu,i} + m_{\nu,j}\right) \cosh(2\Im\omega) \pm \sqrt{(m_{\nu,i} + m_{\nu,j})^2 \cosh^2(\Im\omega) - 4\,m_{\nu,i} m_{\nu,j}} \right] \leq 8\pi\,.
\end{align}

\begin{table}[t!]
    \centering
    \begin{tabular}{c | c c}
        & \multicolumn{2}{c}{$M_N < \dots\,\left[\si{GeV} \right]$} \\[0.5ex]
        & Normal ordering & Inverted ordering \\ \hline
         $J = 0$            & \num{1.30e16} & \num{8.01e15}  \\[1ex]
         $J = \frac{1}{2}$  & \num{1.52e16} & \num{1.59e16} \\[1ex] 
         $J = 1$            & \num{9.33e15} & \num{7.60e15}
    \end{tabular}
    \caption{Maximum allowed value of the HNL masses depending on the scattering processes considered for each neutrino mass ordering.}
    \label{tab:max_values_mass}
\end{table}

The case for $J = 1$ is much more delicate since the matrix to diagonalize in Eq.~\eqref{eq:partial_wave_matrix_J1} becomes a $14 \times 14$ matrix that is much harder to obtain analytically.

As we mentioned earlier, our main interest in the generalized case is to obtain the value of the partial waves when $\omega = 0$ on the seesaw line. This would give us an ``upper bound" on the HNL masses, such that unitarity at tree level is maintained. We can easily obtain analytic solutions for $J = 0$ and $J = \frac{1}{2}$; we have to set $\omega = 0$ for Eqs.~\eqref{eq:bound_general_casasIbarra_J=0} and \eqref{eq:bound_general_casasIbarra_J=1/2} and solve for $M_N$, given a set of neutrino masses. We also provide the results for $J=1$, which we could only obtain numerically. These results are in Table~\ref{tab:max_values_mass}, and in Fig.~\ref{fig:seesaw_line_unitarity} where we compare the more general results with those derived in Section~\ref{sec:scattering_amplitudes}. 

All of the results agree with the general lore that neutrino masses are generated, at most, around the GUT scale ($\sim \SI{e15}{GeV}$) if we want tree-level unitarity to hold \cite{Maltoni:2000iq}.

\section{Generalization beyond two HNLs and the Casas-Ibarra parametrization}
\label{sec:beyond_2HNLs}
As we mentioned before analysis done in Section~\ref{sec:beyond_2HNLs} applies to any model with HNLs with minimal interactions, as long as the Yukawa matrix is rank one. 
That's because choose a basis in the Lagrangian where only one linear combination of HNL states interactions with one linear combination of the lepton doublet. 
We explicitly showed this for the minimal case where we have two degenerate HNLs, as it can be easily seen from having a large value of of $\Im(\omega)$ in the Casas-Ibarra parametrization in Eq.~\eqref{eq:casas_ibarra_matrices}.

Let's conside now the case with three HNLs. 
In the Casas-Ibarra parametrization, the argument that the matrix is approximately rank one is similar to the case for two HNLs. 
Without loss of generality, we can choose the parametrization

\begin{align}
    O = \begin{pmatrix}
        1 & 0 & 0 \\ 
        0 & \cos{\omega_{23}} & \sin{\omega_{23}} \\ 
        0 & -\sin{\omega_{23}} & \cos{\omega_{23}} 
    \end{pmatrix}
    \begin{pmatrix}
        \cos{\omega_{13}} & 0 & \sin{\omega_{13}} \\ 
        0 & 1 & 0 \\
        -\sin{\omega_{13}} & 0 & \cos{\omega_{13}}
    \end{pmatrix}
    \begin{pmatrix}
        \cos{\omega_{12}} & \sin{\omega_{12}} & 0 \\
        - \sin{\omega_{12}} & \cos{\omega_{12}} & 0 \\
        0 & 0 & 1
    \end{pmatrix}\,,
\end{align}
setting $\Im(\omega_{13}) \gg 1$\footnote{Had we chosen any of the other angles to be the larger one, it would've left one of the different HNLs to be completely sterile in Yukawa interactions}, and expanding $O$ in the leading terms of $\Im(\omega_{13})$
\begin{align}
    O \simeq \frac{e^{-i\omega_{13}}}{2} 
    \begin{pmatrix}
        1 & 0 & 0 \\
        0 & c_{23} & s_{23} \\ 
        0 & -s_{23} & c_{23}
    \end{pmatrix}
    \begin{pmatrix}
        1 & 0 & i \\
        0 & 0 & 0 \\
        -i & 0 & 1
    \end{pmatrix}
    \begin{pmatrix}
        c_{12} & s_{12} & 0 \\
        -s_{12} & c_{12} & 0 \\
        0 & 0 & 1
    \end{pmatrix}\,,
\end{align}
where $c_{ij} \equiv \cos(\omega_{ij}), s_{ij} \equiv \sin(\omega_{ij})$. It is clear from this shape that $O$ is a rank one matrix. From this parametrization of the Yukawa matrix, once again, only one specific linear combination of HNLs interact with another specific linear combination of the lepton doublet. In this case, the linear combinations are
\begin{align}
    N_R &= \frac{1}{\sqrt{\abs{c_{12}}^2 + \abs{s_{12}}^2 + 1}} \left(c_{12}\,N_{1,R} + s_{12}\,N_{2,R} + i\,N_{3,R} \right)\,, \\
    \nu_L &= \frac{1}{\sqrt{m_{\nu,1} + m_{\nu,2}\,\abs{s_{23}}^2 + m_{\nu,3}\,\abs{c_{23}}^2}}\left( \sqrt{m_{\nu,1}}\,\nu_{1,L} + i\,\sqrt{m_{\nu,2}}\,s_{23}^\ast\,\nu_{2,L} + i \sqrt{m_{\nu,3}}\,c_{23}^\ast\,\nu_{3,L} \right)\,, \\
    \ell_L &= \frac{1}{\sqrt{m_{\nu,1} + m_{\nu,2}\,\abs{s_{23}}^2 + m_{\nu,3}\,\abs{c_{23}}^2}}\left( \sqrt{m_{\nu,1}}\,\ell_{1,L} + i\,\sqrt{m_{\nu,2}}\,s_{23}^\ast\,\ell_{2,L} + i \sqrt{m_{\nu,3}}\,c_{23}^\ast\,\ell_{3,L} \right)\,,
\end{align}
where $L_{i,L} = \sum_\alpha V_{\alpha i} L_{\alpha, L}$ and therefore all results in Section~\ref{sec:scattering_amplitudes} apply to this case as well, where now 
\begin{align}
\abs{Y_{\mathrm{tot}}}^2 \simeq \frac{g^2}{8\,M_W^2}\,\left(\abs{c_{12}}^2 + \abs{s_{12}}^2 + 1 \right) \left(m_1 + m_2\,\abs{s_{23}}^2 + m_3\,\abs{c_{23}}^2\right) M_N\,e^{2\Im \omega_{12}}\,,
\end{align}

Beyond the Casas-Ibarra parametrization, it will generally be the case that minimal seesaw models with large mixing angles and radiatively stable neutrino masses (or equivalently, an approximate lepton conserving symmetry) to be approximately rank one \cite{Shaposhnikov:2006nn, Abada:2007ux, Kersten:2007vk, Adhikari:2010yt, Moffat:2017feq}. 
For example, let us consider the parametrizations of Yukawa matrices in \cite{Kersten:2007vk}, for both two and three HNLs:
\begin{align}
\label{eq:rank_one_Y}
    Y_{2N} &= \begin{pmatrix}
        Y_1 & Y_2 \\
        \alpha\,Y_1 & \alpha\,Y_2 \\
        \beta\,Y_1 & \beta\,Y_2
    \end{pmatrix}\,, & 
    Y_{3N} &= \begin{pmatrix}
        Y_1 & Y_2 & Y_3 \\
        \alpha\,Y_1 & \alpha\,Y_2 & \alpha\,Y_3  \\
        \beta\,Y_1 & \beta\,Y_2 & \beta\,Y_3
    \end{pmatrix}\,,
\end{align}
in both of these scenarios, as we would expect from rank one Yukawa matrices, only one linear combination of the lepton doublet interacts with one linear combination of HNLs (this is also said and shown explicitly in \cite{Kersten:2007vk}, for this parametrization). Specifically,
\begin{align}
    N_R &= \frac{Y_1\,N_{1,R} + Y_2\,N_{2,R} + Y_3\,N_{3,R}}{\sqrt{\abs{Y_1}^2 + \abs{Y_2}^2 + \abs{Y_3}^2}}\,, \\
    \nu_L &= \frac{\nu_{e,L} + \alpha^\ast \nu_{\mu,L} + \beta^\ast \nu_{\tau,L}}{\sqrt{1 + \abs{\alpha}^2 + \abs{\beta}^2 }}\,, \\
    \ell_L &= \frac{e_L + \alpha^\ast \mu_L + \beta^\ast \tau_L}{\sqrt{1 + \abs{\alpha}^2 + \abs{\beta}^2 }}\,,
\end{align}
for the case with three HNLs (the case of two HNLs is similar).
These rank one matrices will therefore give at most one non-zero neutrino mass. 
Our results (obtained above for the Casas-Ibarra parametrization) apply to the parametrization~\eqref{eq:rank_one_Y} as well.

However, small perturbations from matrix~\eqref{eq:rank_one_Y} generate radiatively stable small neutrino masses \cite{Pilaftsis:1991ug, Kersten:2007vk}, and would make the matrix at least rank two. 
This affects our scattering analysis in the negligible way, proportional to these perturbations. We hypothesize that this should only be relevant close to the seesaw line, where the mixing angles are those from the canonical seesaw $\abs{\Theta}^2 \propto m_\nu/M_N$. 
We performed such an  analysis for the minimal two HNL case in Section~\ref{sec:seesaw_line}.

All of these results can also apply to other models with singlet HNLs of spin $1/2$, as long as the Yukawa matrix is rank one.
For models where HNLs are $\mathrm{SU}(2)_L$ multiplets, such as the Type-III Seesaw \cite{Foot:1988aq} these results may not apply; 
as well as for models where HNLs have spin $3/2$ \cite{Demir:2021cuo} (see also \cite{Li:2022abx}). 
Gauge extensions of the SM, like the Left-Right Symmetric Models \cite{Mohapatra:1979ia} or Grand Unified Theories \cite{Fritzsch:1974nn, Georgi:1974my, delAguila:1980qag}, would require a more involved analysis since they can introduce additional Yukawa interactions with a more complicated scalar sector, and moreover it might also be the case that these gauge extensions already predict a specific shape for the Yukawa matrices.
Our analysis in principle can also apply to Seesaw realizations with more than one energy scale, such as the inverse \cite{Mohapatra:1986bd}, linear \cite{Akhmedov:1995vm, Malinsky:2005bi}, or extended seesaws \cite{Mohapatra:1986aw}. 
However, these models don't require a Yukawa matrix of rank one to have large mixing angles and small neutrino masses, and we can allow ourselves to have more general shapes of Yukawa matrices; a more involved analysis might therefore be needed for these models, similar to the ones presented in Section~\ref{sec:seesaw_line} with the matrices used in Appendix~\ref{app:general_matrices}.

\section{Summary and conclusions} 
\label{sec:conclusions}

The addition of heavy neutral leptons remains one of the most compelling extensions to explain the origin of neutrino masses. However, this framework does not predict the mass of these particles or the magnitude of their mixing with Standard Model neutrinos.

The parameters of heavy neutral leptons can be constrained through tree-level unitarity arguments. While the $S$ matrix is always unitary in any renormalizable theory, apparent violations of unitarity at tree level signal a breakdown of perturbation theory. This loss of perturbativity provides a practical method for setting limits on the masses and couplings of heavy neutral leptons.

In this work, we identify the region of parameter space where the minimal type I seesaw model would cease to be unitary at tree level. We find that unitarity breaks down when $\abs{Y_{\mathrm{tot}}}^2 \geq 8\pi/\varphi \simeq \num{15.533}\dots$, assuming the Yukawa matrix is of rank one.

The Yukawa matrix remains approximately rank one in scenarios with large mixing angles and small neutrino masses. The deviation from rank one is proportional to the neutrino masses and is only relevant when our scattering analysis is performed near the so-called \emph{seesaw line}. An analysis in this regime provides the maximal value for the heavy neutral lepton masses compatible with unitarity and successful neutrino mass generation.
We find that this upper limit lies between $10^{15}$ and $10^{16},\si{GeV}$ for both inverted and normal ordering of neutrino masses, in line with the usual expectation that neutrino masses originate near the GUT scale.

We would also like to state that there is nothing wrong with the values of the theory not satisfying the tree-level unitarity or perturbativity. 
We can always refine the unitarity 
constraints
by doing a one loop analysis on the model, which we would expect would ameliorate the
parameter space allowed
\cite{Logan:2022uus}, or it was suggested by \cite{Passarino:1985ax} it could refine the
inequalities and restrict the parameter space even more.
Higher order corrections will change the behavior of the amplitudes and also of the $S$-matrix and amplitudes. 
It is hard to know exactly how, but previous analyses applied to the Higgs boson and other extended BSM models \cite{Dawson:1988va, Dawson:1989up, Durand:1989zs, Durand:1991yf, Durand:1993vn, Grinstein:2015rtl} give us hints that it might allow for larger couplings such that unitarity is kept at higher orders.
It is even more unclear the results one would get from two-loops.

Of course, there is nothing wrong with the breakdown of perturbation theory.
One of the pillars of the SM, quantum chromodynamics (QCD), violates perturbativity; lattice and several next-to-leading order computations are necessary to generate results that agree with experiments.
No one or nothing is telling us that the theory that generates neutrino masses should not be strongly interacting, a possibility seldom discussed in the literature.

Our results apply to any model with singlet HNLs and minimal interactions, as long as the Yukawa couplings are rank one, or are approximately rank one. 
For other seesaw realizations with HNLs, such as the minimal Type-III seesaw, or other realizations from gauge extensions our results may not apply. 
In extensions where there is more than one scale, such as the inverse, linear, or extended seesaw, in principle these results apply if the parametrization of the Yukawa matrix is rank one. 
However this extensions have small neutrino masses and large mixing angles without the need of a special parametrization of the Yukawa matrix. Therefore this analysis is probing a region of the parameter space which may not be phenomenologically relevant for this models.
A dedicated analysis, in the same vein as our paper, for all these models is therefore needed.

\acknowledgments
We want to thank Poul Henrik Damgaard, Matt von Hippel, Matthias Wilhelm, Mikhail Shaposhnikov, and especially Fedor Bezrukov for their helpful and meaningful discussions. 

This project has received funding from the European Research Council (ERC) under the European Union’s Horizon 2020 research and innovation programme (Prgram No. GA 694896) and from the Carlsberg Foundation (grant agreement CF17-0763).
The work of I.T. was partially supported by the European Union’s Horizon 2020 research and innovation program under the Marie Sklodowska-Curie grant agreement No. 847523 ‘INTERACTIONS’.

\appendix

\section{Definitions and conventions} \label{app:def_and_con}
In this Appendix we write the definitions and conventions of different important functions.

\subsection{Wigner \texorpdfstring{$d$}{Lg}-functions}
We collect here some properties and definitions related to Wigner $d$-functions. For a more elaborate discussion, we refer the reader to Chapter 4 of \cite{Varshalovich:1988ifq}.

The Wigner $D$ functions are defined as matrix elements of the rotation operator $\mathcal{R}(\phi, \theta, \psi) = e^{-i\phi J_x}\, e^{-i \theta J_y}\,e^{-i \psi J_z}$ where $J_x, J_y, J_z$ are the generators of the $\mathrm{SU}(2)$ group. Then
\begin{equation}
    D^j_{m^\prime,m}(\phi, \theta, \psi) \equiv \mel{jm^\prime}{\mathcal{R}(\phi, \theta, \psi)}{jm} = e^{-m^\prime \phi} d_{m^\prime, m}^j(\theta) e^{-i m \psi}\,,
\end{equation}
where $j = 0, \frac{1}{2}, 1, \dots$ and $m = -j, -j+1, \dots, j-1, j$. The Wigner (small) $d$-functions are
\begin{equation}
    d_{m^\prime,m}^{j}(\theta) = D^j_{m^\prime,m}(0, \theta, 0) = \mel{jm^\prime}{e^{-\theta J_y}}{jm}\,,
\end{equation}
where $d_{m^\prime,m}^{j}(\theta)$ are real functions that follow
\begin{align}
    \int_{-1}^1 \dd(\cos\theta)\,d_{m^\prime,m}^j(\theta)\,d_{m^\prime,m}^{j^\prime}(\theta) = \frac{2\,\delta_{jj^\prime}}{2j+1}\,, \\[2ex]
    d_{m^\prime,m}^j(\theta) = d_{-m,-m^{\prime}}^j = (-1)^{m^\prime - m} d_{m,m^{\prime}}^j\,.
\end{align}

The explicit expression for some Wigner $d$-functions used for our analysis are
\begin{align}
    d_{0,0}^0(\theta) &= 1\,,\\[2ex]
    d_{\frac{1}{2},\frac{1}{2}}^{\frac{1}{2}}(\theta) &= \cos\frac{\theta}{2}\,,\\[2ex]
    d_{01}^1(\theta) &= -\frac{\sin\theta}{\sqrt{2}}\,,\hspace{1cm} d_{11}^1(\theta) = \cos^2\frac{\theta}{2}\,,
\end{align}

\subsection{Spinor helicity formalism}
The helicity spinors in the chiral basis are (see Appendix A of \cite{Giunti:2007ry})
\begin{align}
    u_h(p) &= \begin{pmatrix}
        -\sqrt{E + h \abs{\vec{p}}}\,\chi_h(\hat{p}) \\
        \sqrt{E - h \abs{\vec{p}}}\,\chi_h(\hat{p})
    \end{pmatrix}\,, &
    v_h(p) &= -h\,\begin{pmatrix}
        \sqrt{E - h \abs{\vec{p}}}\,\chi_{-h}(\hat{p}) \\
        \sqrt{E + h \abs{\vec{p}}}\,\chi_{-h}(\hat{p})
    \end{pmatrix}\,,
\end{align}
where $h, E$, $\vec{p}$ and $\hat{p}$ are the helicity, energy, three momentum, and direction of three-momentum of the particle in question. The two-component helicity spinors, $\chi_h$, are 
\begin{align}
    \chi_+(\hat{p}) &= \begin{pmatrix}
        \cos\frac{\theta}{2} \\ \sin\frac{\theta}{2}\,e^{i\phi}
    \end{pmatrix}\,, &
    \chi_-(\hat{p}) &= \begin{pmatrix}
        -\sin\frac{\theta}{2}\,e^{-i\phi} \\ \cos\frac{\theta}{2}
    \end{pmatrix}\,,
\end{align}
where $\theta$ and $\phi$ come from the parametrization of the direction of three momentum
\begin{align}
    \hat{p}(\theta, \phi) = \left(\sin\theta\cos\phi, \sin\theta\sin\phi, \cos\theta \right)\,.
\end{align}

For our case, we were dealing exclusively with $2 \to 2$ scatterings with a center of mass energy much larger than any of the masses of the particles. We parametrize the four four-momenta $p_a + p_b \to p_1 + p_2$ as
\begin{align}
    p_a &= \frac{\sqrt{s}}{2}\left(1, \hat{p}(0, \phi) \right)\,,& p_b &= \frac{\sqrt{s}}{2}\left(1, \hat{p}(\pi, \phi + \pi) \right)\,,  \\ 
    p_1 &= \frac{\sqrt{s}}{2}\left(1, \hat{p}(\theta, \phi) \right)\,, & p_2 &= \frac{\sqrt{s}}{2}\left(1, \hat{p}(\pi - \theta, \phi + \pi) \right) \,.
\end{align}

\section{
Results for general Yukawa couplings} \label{app:general_yukawa}
In this Appendix, we will present the amplitudes for the different processes we considered. In the main text, we presented the results in simplified cases with an approximate $L$ symmetry, where we effectively only have one HNL interacting with one lepton doublet.

Without making this consideration, the Lagrangian reads as
\begin{align}
    \label{eq:yukawa_interaction_general}
    \mathcal{L}_Y &= -\sum_{\alpha,i} Y_{\alpha i}\left[\bar{\nu}_{\alpha, L}\,N_{i,R}\,\phi^{0\ast}  - \bar{\ell}_{\alpha, L}\,N_{i,R}\,\phi^-  \right] + \mathrm{H.c.}\,,
\end{align}
where $\alpha \in \left(e,\mu,\tau \right)$ and $i = 1,\dots,\mathcal{N}$ where $\mathcal{N}$ is the number of HNLs we decided to add to our theory.

The Feynman rules of the interactions in Eq.~\eqref{eq:yukawa_interaction_general} should be straightforward as well. 

In the remaining part of the Appendix, we will present the shape of the amplitudes of all the processes we considered without any assumptions of the particular shape of Yukawa particles and then present the shape of partial waves. The results in the main text are recovered if we substitute the combination of Yukawa couplings by $\abs{Y_{\mathrm{tot}}}^2$.

\subsection{\texorpdfstring{$J = 0$}{Lg} helicity amplitudes}
The amplitudes for both processes in Eq.~\eqref{eq:j_0_process_1} and \eqref{eq:j_0_process_2} are the same. They are both due to an $s$ channel. Their amplitudes are
\begin{align}
    \label{eq:j_0_amplitude_general_1}
    &i\mathcal{M}\left(N_i \ell_\alpha^\mp \to N_j \ell_\beta^\mp\right): \hspace{1cm}
    \left\lbrace
    \begin{aligned}
        - - \to - - &= -Y_{\alpha i}^{\ast}\,Y_{\beta j}^{\phantom{\ast}}\,, \\
        + + \to + + &= -Y_{\alpha i}^{\phantom{\ast}}\,Y_{\beta j}^{\ast}\,, \\
    \end{aligned} \right. \\[2ex]
    \label{eq:j_0_amplitude_general_2}
    &i\mathcal{M}\left(N_i \nu_\alpha \to N_j \nu_\beta\right): \hspace{1cm}
    \left\lbrace
    \begin{aligned}
        - - \to - - &= -Y_{\alpha i}^{\ast}\,Y_{\beta j}^{\phantom{\ast}}\,, \\
        + + \to + + &= -Y_{\alpha i}^{\phantom{\ast}}\,Y_{\beta j}^{\ast}\,, \\
    \end{aligned} \right.
\end{align}

\subsection{\texorpdfstring{$J = \frac{1}{2}$}{Lg} helicity amplitudes}
\begin{align}
    \label{eq:j_1/2_amplitude_general_1}
    &i\mathcal{M}\left(N_i \phi^+ \to N_j \phi^+\right): \hspace{1cm}
    \left\lbrace
    \begin{aligned}
        + 0 \to + 0 = -\sum_{\gamma} Y_{\gamma i}^{\phantom{\ast}}\,Y_{\gamma j}^{\ast}\,\sec(\theta/2)\,, \\
        - 0 \to - 0 = -\sum_{\gamma} Y_{\gamma i}^{\phantom{\ast}}\,Y_{\gamma j}^{\ast}\,\cos(\theta/2)\,,
    \end{aligned} \right. \\[2ex]
    \label{eq:j_1/2_amplitude_general_2}
    &i\mathcal{M}\left(N_i \phi^0 \to N_j \phi^0\right): \hspace{1cm}
    \left\lbrace
    \begin{aligned}
        + 0 \to + 0 = -\sum_{\gamma} Y_{\gamma i}^{\phantom{\ast}}\,Y_{\gamma j}^{\ast}\,\sec(\theta/2)\,, \\
        - 0 \to - 0 = -\sum_{\gamma} Y_{\gamma i}^{\phantom{\ast}}\,Y_{\gamma j}^{\ast}\,\cos(\theta/2)\,,
    \end{aligned} \right.
\end{align}
\begin{align}
    \label{eq:j_1/2_amplitude_general_3}
    i\mathcal{M}\left(\ell_{\alpha}^- \phi^+ \to \ell_{\beta}^- \phi^+\right): \hspace{1cm}
        &- 0 \to - 0 = -\sum_{k} Y_{\alpha k}^{\ast}\,Y_{\beta k}^{\phantom{\ast}}\,\cos(\theta/2)\,, \\[2ex]
    \label{eq:j_1/2_amplitude_general_4}
    i\mathcal{M}\left(\ell_{\alpha}^- \phi^+ \to \nu_{\beta} \phi^0\right): \hspace{1cm}
        &- 0 \to - 0 = \sum_{k} Y_{\alpha k}^{\ast}\,Y_{\beta k}^{\phantom{\ast}}\,\cos(\theta/2)\,, \\[2ex]
    \label{eq:j_1/2_amplitude_general_5}
    i\mathcal{M}\left(\nu_{\alpha} \phi^0 \to \nu_{\beta} \phi^0\right): \hspace{1cm}
        &- 0 \to - 0 = -\sum_{k} Y_{\alpha k}^{\ast}\,Y_{\beta k}^{\phantom{\ast}}\,\cos(\theta/2)\,, \\[2ex]
        i\mathcal{M}\left(\ell_{\alpha}^- \phi^- \to \ell_{\beta}^- \phi^-\right): \hspace{1cm}
        &- 0 \to - 0 = -\sum_{k} Y_{\alpha k}^{\ast}\,Y_{\beta k}^{\phantom{\ast}}\,\sec(\theta/2)\,, \\[2ex]
        i\mathcal{M}\left(\nu_{\alpha} \phi^{0\ast} \to \nu_{\beta} \phi^{0\ast}\right): \hspace{1cm}
        &- 0 \to - 0 = -\sum_{k} Y_{\alpha k}^{\ast}\,Y_{\beta k}^{\phantom{\ast}}\,\sec(\theta/2)\,, \\[2ex]
        i\mathcal{M}\left(\nu_\alpha \phi^- \to \ell_{\beta}^- \phi^{0\ast}\right): \hspace{1cm} 
        &- 0 \to - 0 = \sum_{k} Y_{\alpha k}^{\ast}\,Y_{\beta k}^{\phantom{\ast}}\,\sec(\theta/2)\,,
\end{align}

\subsection{\texorpdfstring{$J = 1$}{Lg} helicity amplitudes}
\begin{align}
    &i\mathcal{M}\left(N_i \ell_\alpha^\mp \to N_j \ell_\beta^\mp\right): \hspace{1cm}
    \left\lbrace
    \begin{aligned}
        + - \to + - &= Y_{\alpha j}^{\ast}\,Y_{\beta i}^{\phantom{\ast}}\,, \\
        - + \to - + &= Y_{\alpha j}^{\phantom{\ast}}\,Y_{\beta i}^{\ast}\,, \\
    \end{aligned} \right. \\[2ex]
    &i\mathcal{M}\left(N_i \nu_\alpha \to N_j \nu_\beta\right): \hspace{1cm}
    \left\lbrace
    \begin{aligned}
        + - \to + - &= Y_{\alpha j}^{\ast}\,Y_{\beta i}^{\phantom{\ast}}\,, \\
        - + \to - + &= Y_{\alpha j}^{\phantom{\ast}}\,Y_{\beta i}^{\ast}\,, \\
    \end{aligned} \right. \\[2ex]
     &i\mathcal{M}\left(\nu_\alpha \ell_\beta^- \to \phi^0 \phi^-\right): \hspace{1cm} 
     \phantom{\lbrace}
     + - \to 0\,0 = -\sum_k Y_{\alpha k}^{\phantom{\ast}}\,Y_{\beta k}^{\ast} \cot(\theta/2)\,, \\[2ex]
     &i\mathcal{M}\left(N_i N_j \to \nu_\alpha \nu_\beta\right): \hspace{1cm} 
     \left\lbrace
    \begin{aligned}
        + - \to + - &= Y_{\alpha j}^{\ast}\,Y_{\beta i}^{\phantom{\ast}}\,, \\
        - + \to - + &= Y_{\alpha j}^{\phantom{\ast}}\,Y_{\beta i}^{\ast}\,, \\
    \end{aligned} \right. \\[2ex]
    &i\mathcal{M}\left(N_i N_j \to \ell_\alpha^{\mp} \ell_\beta^{\pm}\right): \hspace{1cm} 
     \left\lbrace
    \begin{aligned}
        + - \to + - &= Y_{\alpha j}^{\ast}\,Y_{\beta i}^{\phantom{\ast}}\,, \\
        - + \to - + &= Y_{\alpha j}^{\phantom{\ast}}\,Y_{\beta i}^{\ast}\,, \\
    \end{aligned} \right. \\[2ex]
    \label{eq:j_1_amplitude_general_1}
     &i\mathcal{M}\left(N_i N_j \to \phi^+ \phi^-\right): \hspace{1cm} 
     \phantom{\lbrace}
     - + \to 0\,0 = -\sum_\gamma Y_{\gamma i}^{\ast}\,Y_{\gamma j}^{\phantom{\ast}}\,\tan(\theta/2)\,, \\[2ex]
     &i\mathcal{M}\left(N_i N_j \to \phi^0 \phi^{0\ast}\right): \hspace{1cm} 
     \phantom{\lbrace}
     - + \to 0\,0 = -\sum_\gamma Y_{\gamma i}^{\ast}\,Y_{\gamma j}^{\phantom{\ast}}\,\tan(\theta/2)\,, \\[2ex]
     &i\mathcal{M}\left(\nu_\alpha \nu_\beta \to \phi^0 \phi^{0\ast}\right): \hspace{1cm} 
     \phantom{\lbrace}
     - + \to 0\,0 = -\sum_k Y_{\alpha k}^{\ast}\,Y_{\beta k}^{\phantom{\ast}}\,\tan(\theta/2)\,, \\[2ex]
     \label{eq:j_1_amplitude_general_2}
     &i\mathcal{M}\left(\ell_\alpha^- \ell_\beta^+ \to \phi^+ \phi^-\right): \hspace{1cm} 
     \phantom{\lbrace}
     - + \to 0\,0 = -\sum_k Y_{\alpha k}^{\ast}\,Y_{\beta k}^{\phantom{\ast}}\,\tan(\theta/2)\,,
\end{align}

\subsection{General shape of partial wave matrices} \label{app:general_matrices}
From Eq.~\eqref{eq:j_0_amplitude_general_1} and \eqref{eq:j_0_amplitude_general_2}, we can derive the $J = 0$ partial wave matrix. We have to consider the fact that if we are dealing with three different flavors and $\mathcal{N}$ different HNLs, then our matrix is a $(3 + \mathcal{N})$ square matrix.

The partial wave matrix is
\begin{align}
    \mathbf{a}^{J = 0} = -\frac{1}{16\pi} \begin{pmatrix}
        \abs{Y_{e1}^{\phantom{\ast}}}^2 & Y_{e1}^\ast\,Y_{\mu 1}^{\phantom{\ast}} & Y_{e 1}^\ast\,Y_{\tau 1}^{\phantom{\ast}} & Y_{e 1}^\ast\,Y_{e 2}^{\phantom{\ast}} & \cdots & Y_{e 1}^\ast\,Y_{\tau \mathcal{N}}^{\phantom{\ast}} \\[1.5ex]
        Y_{\mu 1}^\ast\,Y_{e 1}^{\phantom{\ast}} & \abs{Y_{\mu 1}^{\phantom{\ast}}}^2 & Y_{\mu 1}^\ast\,Y_{\tau 1}^{\phantom{\ast}} & Y_{\mu 1}^\ast\,Y_{e 2}^{\phantom{\ast}} & \cdots & Y_{\mu 1}^\ast\,Y_{\tau \mathcal{N}}^{\phantom{\ast}} \\[1.5ex]
        Y_{\tau 1}^\ast\,Y_{e 1}^{\phantom{\ast}} & Y_{\tau 1}^\ast\,Y_{\mu 1}^{\phantom{\ast}} & \abs{Y_{\tau 1}^{\phantom{\ast}}}^2 & Y_{\tau 1}^\ast\,Y_{e 2}^{\phantom{\ast}} & \cdots & Y_{\tau 1}^\ast\,Y_{\tau \mathcal{N}}^{\phantom{\ast}} \\
        \vdots & \vdots & \vdots & \vdots & \ddots & \vdots \\
        Y_{\tau \mathcal{N}}^\ast\,Y_{e 1}^{\phantom{\ast}} & Y_{\tau \mathcal{N}}^\ast\,Y_{\mu 1}^{\phantom{\ast}} & Y_{\tau \mathcal{N}}^\ast\,Y_{\tau 1}^{\phantom{\ast}} & Y_{\tau \mathcal{N}}^\ast\,Y_{e 2}^{\phantom{\ast}} & \cdots & \abs{Y_{\tau \mathcal{N}}}^2
    \end{pmatrix}\,.
\end{align}
where the initial (final) states of the columns (rows) are
\begin{align}
    \mathbf{a}^{J = 0} = 
    \bordermatrix{
    ~ & N_{1,-}\,e^-_- & N_{1,-}\,\mu^-_- & N_{1,-}\,\tau^-_- & N_{2,-}\,e^-_- & \cdots & N_{\mathcal{N},-}\,\tau^-_- \cr 
    N_{1,-}\,e^-_- \cr
    N_{1,-}\,\mu^-_- \cr
    N_{1,-}\,\tau^-_- \cr
    \hfill\vdots\hfill \cr
    N_{\mathcal{N},-}\,\tau^-_-
    }
\end{align}

The unitarity
inequality apply to each element in the diagonal individually. In order to obtain the best possible
results from this process, we would have to obtain the largest eigenvalue in absolute value. Fortunately, it is easy to do. The matrix is a rank one matrix (meaning it has only one linearly independent row or column), and therefore, we can write it as the product of two vectors
\begin{align}
    \mathbf{a}^{J = 0} = -\frac{1}{16\pi} \begin{pmatrix}
        Y_{e1}^\ast \\ Y_{\mu 1}^\ast \\ \vdots \\ Y_{\tau \mathcal{N}}^\ast
    \end{pmatrix} \begin{pmatrix}
        Y_{e1} & Y_{\mu 1} \cdots Y_{\tau \mathcal{N}}
    \end{pmatrix}\,.
\end{align}
rank one matrices also have the property of having only one non-zero eigenvalue. Then, we can optimize
results by taking the trace
\begin{equation}
    \begin{aligned}
        a^{J=0} = -\frac{1}{16\pi} \sum_{\alpha,i} \abs{Y_{\alpha i}}^2\,, \\[2ex]
    \implies \sum_{\alpha,i} \abs{Y_{\alpha i}}^2 = \abs{Y_{\mathrm{tot}}}^2 \leq 8\pi\,,
    \end{aligned}
\end{equation}
which is exactly the same
results as in Eq.~\eqref{eq:ytot_8pi_bound}. We got this without making any specific assumption on the shape of the Yukawa matrix or the number of additional HNLs. 

The fact that the results in the main text hold in the general case is the exception, rather than the rule. For the rest of the processes, we cannot recover the same results as in the main text.

Taking as an example the processes for $J = \frac{1}{2}$ in Eqs.~\eqref{eq:j_1/2_amplitude_general_1} and \eqref{eq:j_1/2_amplitude_general_2}, both give the same $\mathcal{N} \times \mathcal{N}$ matrix. In the $N_{1,+} \phi^+, N_{2,+} \phi^+, \dots N_{\mathcal{N}},+ \phi^+$ basis is
\begin{align}
    \mathbf{a}^{J=\frac{1}{2}}_{NH} = -\frac{1}{16\pi}\sum_\gamma \begin{pmatrix}
        \abs{Y_{\gamma 1}^{\phantom{\ast}}}^2 & Y_{\gamma 1}^{\phantom{\ast}}\,Y_{\gamma 2}^{\ast} & \cdots & Y_{\gamma 1}^{\phantom{\ast}}\,Y_{\gamma \mathcal{N}}^{\ast} \\[1.5ex]
        Y_{\gamma 2}^{\phantom{\ast}}\,Y_{\gamma 1}^{\ast} & \abs{Y_{\gamma 2}^{\phantom{\ast}}}^2 & \cdots & Y_{\gamma 2}^{\phantom{\ast}}\,Y_{\gamma \mathcal{N}}^{\ast} \\[1.5ex]
        \vdots & \vdots & \ddots & \vdots \\
        Y_{\gamma \mathcal{N}}^{\phantom{\ast}}\,Y_{\gamma 1}^{\ast} & Y_{\gamma \mathcal{N}}^{\phantom{\ast}}\,Y_{\gamma 2}^{\ast} & \cdots & \abs{Y_{\gamma \mathcal{N}}^{\phantom{\ast}}}^2
    \end{pmatrix} = -\frac{1}{16\pi} Y\,Y^\dagger\,.
\end{align}
Here, the matrix does not have a nice and compact shape of its eigenvalues. This matrix is the sum of three rank one matrices. Therefore, it is at most rank-three and has, at most, three non-zero eigenvalues. 

A similar situation arises with the scatterings in Eq.~\eqref{eq:j_1/2_process_3} (or the ones in Eqs.~\eqref{eq:j_1/2_amplitude_general_3}--\eqref{eq:j_1/2_amplitude_general_5}), where we have a $6\times 6$ partial wave matrix. 

In the $e^-_-\,\phi^+, \mu^-_-\,\phi^+, \tau^-_-\,\phi^+, \nu_{e,-}^-\,\phi^0, \nu_{\mu,-}^-\,\phi^0, \nu_{\tau,-}^-\,\phi^0$ the matrix is 
\begin{align}
    \mathbf{a}^{J=\frac{1}{2}}_{LH} = \frac{1}{32\pi}
    \begin{pmatrix}
        -1 & 1 \\ 1 & -1
    \end{pmatrix} \odot
    \sum_{k}\begin{pmatrix}
        \abs{Y_{e k}^{\phantom{\ast}}}^2 & Y_{ek}^\ast\,Y_{\mu k}^{\phantom{\ast}} & Y_{ek}^\ast\,Y_{\tau k}^{\phantom{\ast}} \\[1.5ex]
        Y_{\mu k}^\ast\,Y_{e k}^{\phantom{\ast}} & \abs{Y_{\mu k}^{\phantom{\ast}}}^2 & Y_{\mu k}^\ast\,Y_{\tau k}^{\phantom{\ast}} \\[1.5ex]
        Y_{\tau k}^\ast\,Y_{e k}^{\phantom{\ast}} & Y_{\tau k}^\ast\,Y_{\mu k}^{\phantom{\ast}} & \abs{Y_{\tau k}^{\phantom{\ast}}}^2
    \end{pmatrix}\,,
\end{align}
where $\odot$ denotes the Kronecker product of two matrices. The eigenvalues of the Kronecker product of two matrices are the product of their eigenvalues (see, for example, Chapter 5 of \cite{Merris:1997book}). Our problem reduces to get the eigenvalues of
\begin{align}
    \mathbf{a}^{J=\frac{1}{2}}_{LH} = -\frac{1}{16\pi}
    \sum_{k}\begin{pmatrix}
        \abs{Y_{e k}^{\phantom{\ast}}}^2 & Y_{ek}^\ast\,Y_{\mu k}^{\phantom{\ast}} & Y_{ek}^\ast\,Y_{\tau k}^{\phantom{\ast}} \\[1.5ex]
        Y_{\mu k}^\ast\,Y_{e k}^{\phantom{\ast}} & \abs{Y_{\mu k}^{\phantom{\ast}}}^2 & Y_{\mu k}^\ast\,Y_{\tau k}^{\phantom{\ast}} \\[1.5ex]
        Y_{\tau k}^\ast\,Y_{e k}^{\phantom{\ast}} & Y_{\tau k}^\ast\,Y_{\mu k}^{\phantom{\ast}} & \abs{Y_{\tau k}^{\phantom{\ast}}}^2
    \end{pmatrix} = -\frac{1}{16\pi} Y^\dagger\,Y\,,
\end{align}
which has at most three non-zero eigenvalues. The eigenvalues of an arbitrary Yukawa matrix do not have a compact form.

Finally, as a last example, let us look at the set of scatterings in Eq.~\eqref{eq:j_1_process_4} (or the scatterings in Eq.~\eqref{eq:j_1_amplitude_general_1} -- \eqref{eq:j_1_amplitude_general_2}). Here, the shape of the generalized form of Eq.~\eqref{eq:partial_wave_matrix_J1} is much more complicated than the ones shown previously in this Appendix. 

In a theory with $\mathcal{N}$ HNLs the matrix has dimensions $(\mathcal{N}^2 + 20) \times (\mathcal{N}^2 + 20)$ and is
\begin{align}
    \mathbf{a}^{J = 1}_{NLH} = \frac{1}{32\pi} \begin{pmatrix}
             0 & \mathbf{Y} & \mathbf{Y} & -\sqrt{2}\,\mathscr{Y} & -\sqrt{2}\, \mathscr{Y} \\
             \mathbf{Y}^\dagger & 0 & 0 & -\sqrt{2}\,\tilde{\mathscr{Y}} & 0 \\
             \mathbf{Y}^\dagger & 0 & 0 & 0 & -\sqrt{2}\,\tilde{\mathscr{Y}} \\
             -\sqrt{2}\,\mathscr{Y}^\dagger & -\sqrt{2}\,\tilde{\mathscr{Y}}^\dagger & 0 & 0 & 0 \\
             -\sqrt{2}\,\mathscr{Y}^\dagger & 0 & -\sqrt{2}\,\tilde{\mathscr{Y}}^\dagger & 0 & 0
         \end{pmatrix}\,,
\end{align}
where $\mathbf{Y}$ is a matrix of dimension $\mathcal{N}^2 \times 9$, $\mathscr{Y}$ is a $\mathcal{N}^2 \times 1$ matrix, and $\tilde{\mathscr{Y}}$ is a $9 \times 1$ one. Their shapes are
\begin{align}
    \mathbf{Y} = 
    \begin{pmatrix}
        \abs{Y_{e1}^{\phantom{\ast}}}^2 & Y_{e1}^{\phantom{\ast}}\,Y_{\mu 1}^\ast & Y_{e 1}^{\phantom{\ast}}\,Y_{\tau 1}^\ast &
        \abs{Y_{\mu 1}^{\phantom{\ast}}}^2 & Y_{\mu 1}^{\phantom{\ast}}\,Y_{e 1}^\ast & Y_{\mu 1}^{\phantom{\ast}}\,Y_{\tau 1}^\ast & 
        \abs{Y_{\tau 1}^{\phantom{\ast}}}^2 & Y_{\tau 1}^{\phantom{\ast}}\,Y_{e 1}^\ast & Y_{\tau 1}^{\phantom{\ast}}\,Y_{\mu 1}^\ast \\
        \vdots & \vdots & \vdots & \vdots & \vdots & \vdots & \vdots & \vdots & \vdots \\
        Y_{e \mathcal{N}}^{\phantom{\ast}}\,Y_{e1}^\ast & Y_{e \mathcal{N}}^{\phantom{\ast}}\,Y_{\mu 1}^\ast & Y_{e \mathcal{N}}^{\phantom{\ast}}\,Y_{\tau 1}^\ast & 
        Y_{\mu \mathcal{N}}^{\phantom{\ast}}\,Y_{\mu 1}^\ast 
        & \cdots & \cdots & \cdots & \cdots & \cdots \\[1ex]
        Y_{e 1}^{\phantom{\ast}}\,Y_{e2}^\ast & Y_{e 1}^{\phantom{\ast}}\,Y_{\mu 2}^\ast & Y_{e 1}^{\phantom{\ast}}\,Y_{\tau 2}^\ast & 
        Y_{\mu 1}^{\phantom{\ast}}\,Y_{\mu 2}^\ast
        & \cdots & \cdots & \cdots & \cdots & \cdots \\[1ex]
        \vdots & \vdots & \vdots & \vdots & \vdots & \vdots & \vdots & \vdots & \vdots \\
        \abs{Y_{e \mathcal{N}}^{\phantom{\ast}}}^2 & Y_{e \mathcal{N}}^{\phantom{\ast}}\,Y_{\mu \mathcal{N}}^\ast & Y_{e \mathcal{N}}^{\phantom{\ast}}\,Y_{\tau \mathcal{N}}^\ast &
        \abs{Y_{\mu \mathcal{N}}^{\phantom{\ast}}}^2 
        & \cdots & \cdots & \cdots & \cdots & \cdots
    \end{pmatrix}
\end{align}

\begin{align}
    \mathscr{Y} &= \sum_{\alpha}\begin{pmatrix}
            \abs{Y_{\alpha 1}}^2 \\
            \vdots \\
            Y_{\alpha \mathcal{N}}\,Y_{\alpha 1}^\ast \\
            Y_{\alpha 1}\,Y_{\alpha 2}^\ast \\
            \vdots \\
            \abs{Y_{\alpha \mathcal{N}}}^2 \\
        \end{pmatrix}\,, & 
        \tilde{\mathscr{Y}} &= \sum_{i} \begin{pmatrix}
            \abs{Y_{e i}}^2 \\
            Y_{\mu i}\,Y_{e i}^\ast \\
            Y_{\tau i}\,Y_{e i}^\ast \\
            \abs{Y_{\mu i}}^2 \\
            Y_{e i}\,Y_{\mu i}^\ast \\
            Y_{\tau i}\,Y_{\mu i}^\ast \\
            \abs{Y_{\tau i}}^2 \\
            Y_{e i}\,Y_{\tau i}^\ast \\
            Y_{\mu i}\,Y_{\tau i}^\ast \\
        \end{pmatrix}\,.
\end{align}

The final task would be to obtain the eigenvalues of this matrix. The analytical computation of these eigenvalues is beyond the scope of this paper.

\section{Unitarity beyond \texorpdfstring{$s \to \infty$}{Lg}}
\begin{figure}[ht]
    \centering
    \includegraphics[width = \linewidth]{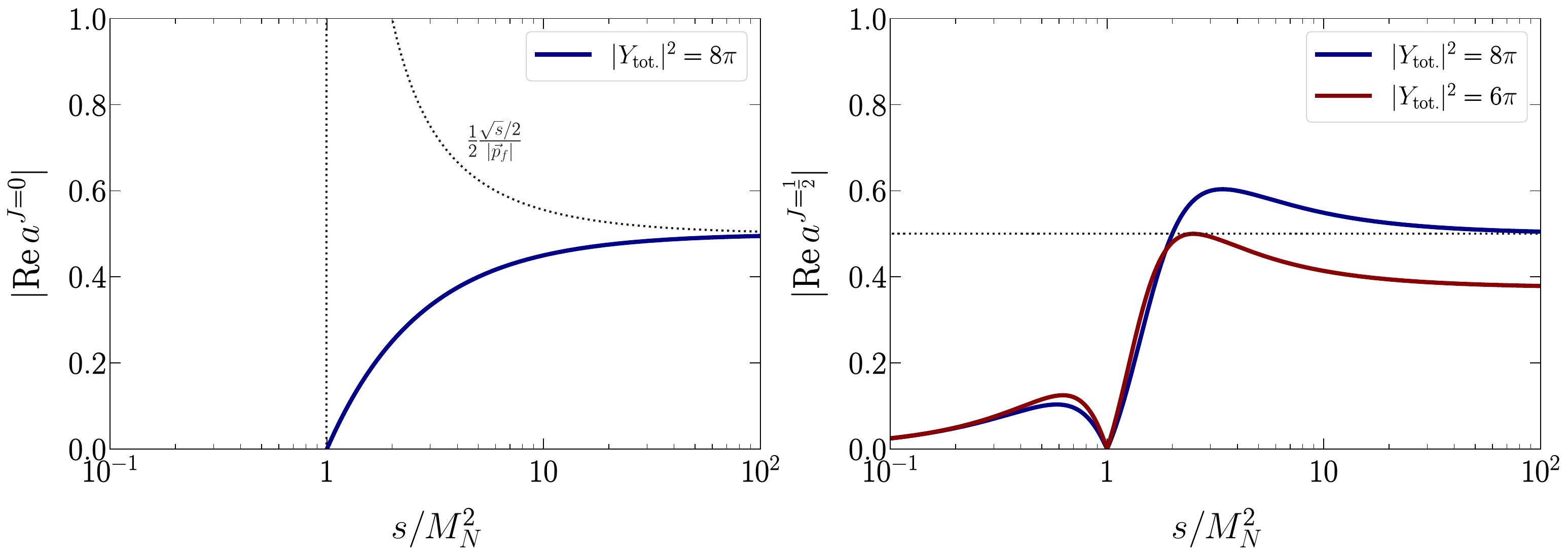}
    \caption{Dependence of the partial waves on $s/M_N^2$, the gray dashed lines on both plots indicate the upper bound that the partial wave cannot be greater than. On the left, the $J = 0$ partial wave changes and the bound (read the text for details). On the right, the $J = \frac{1}{2}$ partial wave changes due to the decay width of the intermediate HNL.}
    \label{fig:massive_plots}
\end{figure}

Throughout the main text, we worked only with amplitudes in the ultra-high energy limit, where the center of mass energy, $\sqrt{s}$, is much bigger than any energy scale in the processes considered.

In this Appendix, we will briefly go through the same analysis we did in the main text with a non-negligible HNL mass. We only examine the cases in $J = 0$, where we have HNLs in the initial and final states, and the $J = \frac{1}{2}$ processes considered in Eqs.~\eqref{eq:j_1/2_process_3}, which are much more interesting because we have an intermediate HNL in the $s$ channel which can be in resonance.

For the $J = 0$ case, given how we have a massive particle in the final state, the inequality relations on Eq.~\eqref{eq:inequalities_for_partial_waves} have to be modified. This is because the integration done in Eq.~\eqref{eq:generalized_optical_theorem} changes if we have massive particles in the final states. The generalized form of Eq.~\eqref{eq:inequalities_for_partial_waves} depends on the three momentum of the final particles, $\abs{\vec{p}_f}$, and is
\begin{align}
    \abs{a_{ii}} &\leq \frac{\sqrt{s}/2}{\abs{\vec{p}_f}}\,, & 0 \leq \Im[a_{ii}] &\leq \frac{\sqrt{s}/2}{\abs{\vec{p}_f}}\,, & \abs{\Re[a_{ii}]} &\leq \frac{1}{2}\,\frac{\sqrt{s}/2}{\abs{\vec{p}_f}}\,.
\end{align}

For all the processes considered in the main text, we have that $\abs{\vec{p}_f} = \sqrt{s}/2$, which is true in the limit where all particles are massless. 

If we were to consider the HNL masses for the processes in Eqs.~\eqref{eq:j_0_process_1} and \eqref{eq:j_0_process_2}, then $\abs{\vec{p}_f} = \frac{s - M_N^2}{2 \sqrt{s}}$. The amplitudes and the partial waves are also modified to
\begin{align}
    a^{J=0} = -&\frac{\abs{Y_{\mathrm{tot}}}^2}{16 \pi}\left(\frac{s - M_N^2}{s}\right) \nonumber \\[1.5ex]
    \implies &\frac{\abs{Y_{\mathrm{tot}}}^2}{16 \pi}\left(\frac{s - M_N^2}{s}\right) \leq \frac{1}{2}\,\frac{s}{s - M_N^2}\,. \label{eq:inequality_J0_mass}
\end{align}
We plot both the left-hand side and the right-hand side of Eq.~\eqref{eq:inequality_J0_mass} in Fig.~\ref{fig:massive_plots} and we find that the
inequality $\mathrm{\abs{Y_{\mathrm{tot}}}^2} \leq 8\pi$ holds for all energies. We can expect the partial wave of any process with an HNL in the initial or final state to have a similar behavior, and the the
strongest constraint would come from $s \to \infty$.

For the $J = \frac{1}{2}$ amplitudes in Eqs.~\eqref{eq:j_1/2_process_3}, we only included the decay width in the renormalized propagator of the intermediate HNL in the $s$ channel. The decay width is 
\begin{align}
    \label{eq:decay_width}
    \Gamma_N = \frac{\abs{Y_{\mathrm{tot}}}^2}{8\pi} M_N\,,
\end{align}
then the partial wave matrix in Eq.~\eqref{eq:j_1/2_partial_waves_matrix_2} gets modified to be
\begin{align}
    \mathbf{a}^{J = 1/2}_{LH} = &\frac{\abs{Y_{\mathrm{tot}}}^2}{32\pi}\left(\frac{s}{s - M_N^2 + i \Gamma_N\,m_N}\right) \begin{pmatrix}
        -1 & 1 \\ 1 & -1
    \end{pmatrix}\,, \\[2ex]
    \label{eq:inequality_bound_J_1/2_mass}
    \implies &\frac{\abs{Y_{\mathrm{tot}}}^2}{16\pi} \abs{\frac{s(s - M_N^2)}{(s - M_N^2)^2 + \Gamma_N^2\,M_N^2}} \leq \frac{1}{2}\,,
\end{align}
with the decay width in Eq.~\eqref{eq:decay_width}, we have that if we want the inequality to hold for all values of $s$ and $M_N^2$, then the unitarity
constraint gets saturated to $\abs{Y_{\mathrm{tot}}}^2 \leq 6\,\pi$. We plot the left hand side of Eq.~\eqref{eq:inequality_bound_J_1/2_mass} for $\abs{Y_{\mathrm{tot}}}^2 = \left\lbrace 6\pi, 8\pi \right\rbrace$ in Fig.~\ref{fig:massive_plots}.

The case for $J = 1$ is much more complicated. After diagonalization, it would be unclear what the final three momentum is, given how we would be dealing with processes where the final particles would have different masses and also different final momentum. Therefore, we will not tackle the problem in this paper. 

\bibliographystyle{JHEP}
\bibliography{bibliography.bib}
\end{document}